%% file: main.tex
\documentclass[camera,letterpaper,nomarginnotes,nonarrowgutter]{jpaper}
\usepackage{mathptmx} 
\usepackage{courier}
\usepackage[italic]{mathastext}

\usepackage[dvipsnames]{xcolor}
\usepackage{fancyhdr}
\usepackage[normalem]{ulem}
\usepackage[sort]{cite}
\usepackage{siunitx}
\usepackage{tikz}
\usepackage{multirow}
\usepackage{enumitem}
\usepackage{algorithm}
\usepackage{listings}
\usepackage{amsmath}
\usepackage{arydshln} 
\usepackage{xspace}
\usepackage{soul} 
\usepackage[bookmarks=true,breaklinks=true,hidelinks]{hyperref}
\usepackage{amssymb}
\usepackage{pifont}
\usepackage{setspace}
\usepackage{footmisc}
\usepackage{balance}
\usepackage{hhline} 
\usepackage[all]{nowidow}

\usepackage[compact]{titlesec}
\usepackage{footnote}
\usepackage{xfrac}

\titlespacing\section{0pt}{2pt plus 0.5pt minus 0.5pt}{-1pt plus 0.5pt minus 0pt}
\titlespacing\subsection{0pt}{2pt plus 0.5pt minus 1pt}{0pt plus 0.5pt minus 0pt}
\titlespacing\subsubsection{0pt}{2pt plus 0.5pt minus 1pt}{2pt plus 0.5pt minus 1pt}

\newcommand{\tech}{IChannels\xspace}
\newcommand{\ta}{IccThreadCovert\xspace}
\newcommand{\tb}{IccSMTcovert\xspace}
\newcommand{\tc}{IccCoresCovert\xspace}


\newif\ifindent
\indentfalse

\ifindent 

\else 

\fi

\newif\ifgraphene
\graphenetrue

\newif\ifcameraready
\camerareadyfalse

\newif\ifonurready
\onurreadyfalse

\newif\ifcommentson
\commentsontrue

\newcommand{\cmark}{\ding{51}}%
\newcommand{\xmark}{\ding{55}}%
\definecolor{darkamber}{rgb}{0.5, 0.19, 0.0}
\definecolor{amber}{rgb}{1.0, 0.49, 0.0}
\definecolor{dkgreen}{rgb}{0,0.6,0}
\definecolor{gray}{rgb}{0.5,0.5,0.5}
\definecolor{mauve}{rgb}{0.58,0,0.82}
\definecolor{lightmauve}{rgb}{0.68,0.4,0.92}
\definecolor{chocolate}{rgb}{0.48, 0.25, 0.0}
\definecolor{dollarbill}{rgb}{0.52,0.73,0.4}
\definecolor{dkdkgreen}{rgb}{0,0.45,0}
\definecolor{gfored}{rgb}{0.580, 0.050, 0.211}
\definecolor{darkwarmgray}{rgb}{0.15, 0.050, 0.05}
\definecolor{ups-truck}{rgb}{0.53, 0.28, 0.21}
\definecolor{azure(colorwheel)}{rgb}{0.0, 0.5, 1.0}

\newcommand{\crj}[1]{\textcolor{black}{#1}}
\newcommand{\cri}[1]{\textcolor{black}{#1}}
\newcommand{\cj}[1]{\textcolor{black}{#1}}
\newcommand{\ja}[1]{\textcolor{black}{#1}}
\newcommand{\jb}[1]{\textcolor{black}{#1}}
\newcommand{\jc}[1]{\textcolor{black}{#1}}
\newcommand{\jd}[1]{\textcolor{black}{#1}}
\newcommand{\je}[1]{\textcolor{black}{#1}}

\newcommand{\jk}[1]{{\color{black}#1}}
\newcommand{\jkv}[1]{{\color{black}#1}}
\newcommand{\jkx}[1]{{\color{black}#1}}

\newcommand{\agythree}[1]{{\color{black}#1}}
\newcommand{\agyfour}[1]{{\color{black}#1}}
\newcommand{\agyc}[1]{\todo[size=\tiny,color=fulvous]{Giray: #1}}
\newcommand{\juang}[1]{{\color{black}#1}}

\ifcommentson
    \newcommand{\agycomment}[1]{}
    \newcommand{\agycrcomment}[1]{\textcolor{dkgreen}{\textbf{[@gy:} #1\textbf{]}}}
    \newcommand{\agycinline}[1]{{\textcolor{dkgreen}{\footnotesize\textbf{[@gy:} #1\textbf{]}}}}
    \newcommand{\agytdinline}[1]{{\textcolor{amber}{\footnotesize\textbf{[TODO:} #1\textbf{]}}}}
    \newcommand{\mpc}[1]{\todo[size=\tiny,color=lightmauve]{#1}}
    \newcommand{\jkc}[1]{\todo[size=\tiny,color=fulvous]{jk: #1}}
    \newcommand{\sgc}[1]{\todo[size=\tiny,color=cyan]{SG: #1}}
    \newcommand{\omc}[1]{\todo[size=\tiny,color=gfored]{\textbf{Onur says:} #1}}
    \newcommand{\loiscomment}[1]{\textcolor{magenta}{\textbf{[@lois:} #1\textbf{]}}}

    \newcommand{\fw}[1]{}
    \newcommand{\atbc}[1]{\todo[size=\tiny,color=ups-truck]{#1}}
\else
    \newcommand{\agycomment}[1]{}
    \newcommand{\agycrcomment}[1]{}
    \newcommand{\loiscomment}[1]{}
    \newcommand{\agycinline}[1]{}
    \newcommand{\agyc}[1]{}
    \newcommand{\atbc}[1]{}
    \newcommand{\agytdinline}[1]{}
    \newcommand{\mpc}[1]{}
    \newcommand{\jkc}[1]{}
    \newcommand{\sgc}[1]{}
    \newcommand{\omc}[1]{}
    \newcommand{\fw}[1]{}
\fi

\ifcameraready
    \newcommand{\agy}[1]{{#1}}
    \newcommand{\agycr}[1]{{#1}}
    \newcommand{\agytwo}[1]{{#1}}
    
    \newcommand{\om}[1]{{#1}}

    \newcommand{\agyrm}[1]{}

    \newcommand{\jkold}[1]{{#1}}
    \newcommand{\lois}[1]{{#1}}

\else 
    \ifonurready
        \newcommand{\agy}[1]{{#1}}
        \newcommand{\agycr}[1]{\textcolor{black}{#1}}
        \newcommand{\agytwo}[1]{\textcolor{black}{#1}}
        \newcommand{\om}[1]{\textcolor{black}{#1}}

        \newcommand{\agyrm}[1]{}

        \newcommand{\jk}[1]{\textcolor{black}{#1}}
        \newcommand{\jkold}[1]{{#1}}
        \newcommand{\lois}[1]{\textcolor{black}{#1}}

    \else
        \newcommand{\agy}[1]{{#1}}
        \newcommand{\agycr}[1]{\textcolor{black}{#1}}
        \newcommand{\agytwo}[1]{{#1}}
        \newcommand{\om}[1]{{#1}}

        \newcommand{\jkold}[1]{{#1}}
        \newcommand{\lois}[1]{\textcolor{black}{#1}}

    \fi 
\fi

\definecolor{burntorange}{rgb}{0.8, 0.33, 0.0}
\definecolor{cadmiumorange}{rgb}{0.93, 0.53, 0.18}
\definecolor{darkorange}{rgb}{1.0, 0.55, 0.0}
\definecolor{fulvous}{rgb}{0.86, 0.52, 0.0}
\definecolor{alizarin}{rgb}{0.82, 0.1, 0.26}
\definecolor{amber}{rgb}{1.0, 0.49, 0.0}

\def\UrlBreaks{\do\/\do-\/\do.\/\do:}

\expandafter\def\expandafter\UrlBreaks\expandafter{\UrlBreaks
  \do\a\do\b\do\c\do\d\do\e\do\f\do\g\do\h\do\i\do\j
  \do\k\do\l\do\m\do\n\do\o\do\p\do\q\do\r\do\s\do\t
  \do\u\do\v\do\w\do\x\do\y\do\z\do\A\do\B\do\C\do\D
  \do\E\do\F\do\G\do\H\do\I\do\J\do\K\do\L\do\M\do\N
  \do\O\do\P\do\Q\do\R\do\S\do\T\do\U\do\V\do\W\do\X
  \do\Y\do\Z}

\newcommand{\squishlist}{
 \begin{list}{$\circ$}
  { \setlength{\itemsep}{0pt}
     \setlength{\parsep}{0pt}
     \setlength{\topsep}{0pt}
     \setlength{\partopsep}{0pt}
     \setlength{\leftmargin}{1em}
     \setlength{\labelwidth}{1em}
     \setlength{\labelsep}{0.5em} } }

\newcommand{\squishsublist}{
\begin{list}{$\rightarrow$}
 { \setlength{\itemsep}{0pt}
    \setlength{\parsep}{0pt}
    \setlength{\topsep}{-10em}
    \setlength{\partopsep}{-3pt}
    \setlength{\leftmargin}{1em}
    \setlength{\labelwidth}{1em}
    \setlength{\labelsep}{0.5em} } }

\newcommand{\squishend}{
  \end{list}  }

\ifcameraready
\else
\fi

\fancypagestyle{firstpage}{

  \pagenumbering{arabic}
}

\newcommand{\affilETH}[0]{\small {}}

\title{\tech: Exploiting Current Management Mechanisms\\to Create Covert Channels in Modern Processors}
\author{
\fontsize{12}{12}\selectfont%
{Jawad Haj-Yahya\affilETH{}}\quad\quad%
{Jeremie S. Kim\affilETH{}}\quad\quad%
{A. Giray Ya\u{g}l{\i}k\c{c}{\i}\affilETH{}}\quad\quad%
{Ivan Puddu\affilETH{}}\quad\quad%
\vspace{0pt}\\%
\fontsize{12}{12}\selectfont%
{Lois Orosa\affilETH{}}\quad\quad%
{Juan Gómez Luna\affilETH{}}\quad\quad%
{Mohammed Alser\affilETH{}}\quad\quad%
{Onur Mutlu\affilETH{}}%
\vspace{1pt}\\%
{\fontsize{11}{11}\selectfont
\affilETH\emph{ETH Z{\"u}rich}%
}
}

 \newcommand{\jr}[1]{{\color{black}{#1}}}

\definecolor{dgreen}{rgb}{0.00, 0.75, 0.00}
\newcommand{\juan}[1]{\textcolor{black}{#1}} 

\newcommand\jkt[1]{{\color{black}{#1}}}

\newcommand{\sr}[1]{{\color{black}{#1}}}

\definecolor{cerulean}{rgb}{0.0, 0.48, 0.65}

\newcommand{\jh}[1]{{\color{black}{#1}}}

\definecolor{schrift}{RGB}{0,73,174}

\newcommand{\hy}[2][yellow]{{%
{#2}}%
}
\newcommand{\hp}[2][pink]{{%
{#2}}%
}
\newcommand{\ho}[2][orange]{{%
{#2}}%
}

\newcommand{\hg}[2][green]{{%
{#2}}%
}

\begin{document}
\bstctlcite{IEEEexample:BSTcontrol}
\maketitle
\ifcameraready
    \thispagestyle{plain}
    \pagestyle{plain}
\else
    \thispagestyle{firstpage}
    \pagestyle{plain}
\fi

\input{_00_abstract}
\input{_01_introduction}
\input{_02_background}
\input{_03_limitations}

\input{_04_technique}

\input{_05_motivation}

\input{_06_results}
\input{_07_mitigation}

\input{_08_related_works}
\input{_09_conclusion}
\section*{Acknowledgments} 
We thank the anonymous reviewers of \agyfour{ASPLOS 2021 and} ISCA 2021 for {feedback}. We thank the
SAFARI Research Group members for {valuable} feedback and the stimulating intellectual environment they provide. We acknowledge the generous gifts provided by our industrial partners: Google, Huawei, Intel, Microsoft, and VMware.

\bibliographystyle{IEEEtranS}
\bibliography{ref}

\end{document}

%% file: _00_abstract.tex

\begin{abstract}
To operate efficiently across a wide range of workloads with varying power requirements, a modern processor applies \sr{different} current management \agy{mechanisms}, which briefly throttle \ja{instruction} execution while they adjust voltage and frequency to accommodate for power-hungry instructions (\ja{PHIs}) in the instruction stream. Doing so 
1) reduces the power consumption of non-PHI instructions in typical workloads and 
2) optimizes system voltage regulators' cost and area for the common use case while limiting current consumption when executing \ja{PHIs}. 

However, these \agy{mechanisms} may compromise \agycr{a system's} confidentiality guarantees\agycr{.} 
In particular, we observe that multi-level side-effects of throttling mechanisms, due to PHI-related current management \agy{mechanisms}, can be detected by two different software contexts \crj{(i.e., sender and receiver)} running on 1) the same hardware \juan{thread}, 2) co-located \ja{Simultaneous Multi-Threading} (SMT) threads, and 3) different physical cores. 

\sr{Based on these} \crj{new} observations on current management \agy{mechanisms}, we develop a new set of covert channels, \sr{\emph{\tech}}, and demonstrate them in real modern \jr{Intel processors} \jr{\agycr{\crj{(}which span more than} $70\%$ of the entire client and server processor market}\agycr{\crj{)}. Our analysis} show\agycr{s} that \tech provides \cj{more than} $24\times$ the channel capacity of state-of-the-art power management covert channels. 
\ja{We} propose practical and effective mitigations to each covert channel in {\tech} \agy{by leveraging the insights we gain through} a rigorous characterization of real systems.

\end{abstract}

%% file: _01_introduction.tex
\section{Introduction}
\label{sec:intro}

Modern high-performance processors support instructions with \jd{widely} varying degrees of \jc{\emph{computational intensity}, due to different instruction types} (e.g., add, multiply, fused-multiply-add\jd{, integer, floating point, vector}) and data widths (\jd{from $32$ to $512$} bits).\jd{\footnote{\jd{We \je{loosely} define computational intensity  as the amount of logic and wire resources an instruction would \je{use} to perform the computation it specifies.}}} This large range of \jc{computational intensity} \jd{across different instructions} results in an instruction set with a wide range of power (current) requirements. For example, \jd{a} \jc{power-hungry instruction (PHI)} \jd{that performs a} 256-bit fused-multiply–add (FMA256) \jc{can} consume approximately $100\times$ the power of a 32-bit register move (MOV32) instruction \cite{haj2015compiler}. This wide range of power requirements feeds into the  many design \om{constraints} of modern processors (e.g., thermal limits, electrical limits, energy consumption)\crj{,} resulting in current management \agy{mechanisms} that allow \crj{the processor} to be energy efficient while dynamically meeting the performance demand.

Current management \agy{mechanisms} dynamically adjust the processor's voltage and frequency based on the power \crj{consumption} of the instructions \crj{that are} being executed while making sure \agytwo{that} design constraints are not violated. 
For example, 256-bit AVX2 instructions \cite{hammarlund2014haswell}  
exhibit significantly larger supply voltage fluctuations  (\emph{di/dt})\crj{\footnote{\jh{Supply voltage fluctuations, known as the \emph{di/dt}, occur when the processor demands rapid changes in load current over a relatively small time scale, due to \crj{large} parasitic inductance in power delivery~\cite{joseph2003control,gough2015cpu,haj2018power,haj2015compiler,larsson1997di}}.}}
compared to \om{lower} power instructions (e.g., \crj{64-bit} scalar, 128-bit SSE)\agycr{, which can lead to voltage drops}. To prevent the voltage from dropping below the minimum operating voltage \agycr{limit} (\ja{$Vcc_{min}$}), the processor increases the voltage guardband of the core before running AVX2 instructions~\cite{haj2015compiler,fetzer2015managing,haj2020flexwatts}.
However, changing the voltage/frequency of the processor can take several microseconds~\ja{\cite{grochowski2004best,haj2018power,hackenberg2015energy,gough2015cpu,huang2015measurement,mazouz2014evaluation}}, whereas a program can change from lower-power instructions to \jc{PHIs} within a few \agycr{clock} cycles (i.e., a few nanoseconds, \emph{three} orders of magnitude lower \crj{latency} than that needed to change the voltage).
To address this disparity, current management \agy{mechanisms} throttle the instruction stream to limit the power (current) of the instructions while adjustments to the voltage/frequency are being made.
As the throttling side-effects of these \agy{mechanisms} are observable across different software contexts, they \jh{can be} used to break the confidentiality guarantees of the whole system.
Recent works~\cite{schwarz2019netspectre, kalmbach2020turbocc} demonstrate methods for creating covert channels via side-effects that arise due to the \jh{two different} throttling \agy{techniques} \om{employed by} the current management \agy{mechanisms} of modern processors: 
\agytwo{1) the \emph{voltage \crj{emergency} avoidance} mechanism~\ja{\cite{joseph2003control,gough2015cpu,haj2018power,reddi2010voltage,reddi2009voltage}} and 2) the \emph{voltage and current limit protection} mechanism~\ja{\cite{ma2014maximum,gough2015cpu,haj2018power,haj2015compiler,varma2015power,2_burton2014fivr,skylakex,intel_avp_2009,naffziger2016integrated,wright2006characterization,piguet2005low,zhang2014architecture}}, both of which we explain in Section~\ref{sec:background}.}
\agytwo{Unfortunately, these recent works are limited in creating covert channels and proposing countermeasures due to the \ja{limited or} inaccurate observations they are built on, as we describe in Section~\ref{sec:limitations}.}

\jk{Our \textbf{goal} is to develop a thorough understanding of \sr{current management} \agy{mechanisms} by rigorously characterizing real modern systems. This allows us to gain several deep insights into how these mechanisms can be abused by attackers, and \crj{how to prevent such vulnerabilities}. \crj{Our experimental} characterization \crj{yields} three \crj{new} observations} about the throttling side-effects of current management \agy{mechanisms}. These observations enable high capacity covert channels between \cri{otherwise isolated \emph{execution contexts}: that is, between different processes, threads, and in general logical partitions of an application with different privilege \ja{levels} (such as \jkv{browser tabs})}. \cri{These covert channels can be established even if the communicating execution contexts are located} 1) on the same hardware thread, 2) \crj{across co-located} \ja{Simultaneous Multi-Threading} (SMT) threads, and 3) \crj{across} different physical cores. \jr{We demonstrate these throttling side-effects in real modern \jr{Intel processors}\agycr{, which span more than} \jr{$70\%$ of the entire client and server processor market \cite{intel_amd_marketshare}.}} \crj{We} refer to \crj{our three new} observations respectively as Multi-Throttling-Thread, Multi-Throttling-SMT, and Multi-Throttling-Cores.


\noindent \textbf{\crj{Observation 1:} Multi-Throttling-Thread.} We find that executing PHIs results in a multi-level reduction of the instructions per cycle (IPC) \ja{performance of} the core.\ja{\footnote{\ja{Instruction supply rate into the backend of the core gets reduced to one of} several discretized levels.}} Effectively, the IPC reduction manifests as a multi-level throttling period length proportional to the instructions' computational intensity. For example, executing a 256-bit vector bit-wise OR instruction (e.g., VORPD-256) results in a throttling period shorter than executing a 512-bit vector multiplication instruction (e.g., VMULPD-512).
\crj{A longer} throttling period is primarily due to the higher voltage guardband requirement of instructions with higher computational intensity, which requires more time \agycr{for the voltage regulator} to ramp up the voltage level\agycr{.}

\noindent \textbf{\crj{Observation 2:} Multi-Throttling-SMT.} We find that in a processor with \ja{Simultaneous Multi-Threading} (SMT)\crj{\cite{tullsen1995simultaneous},} when a thread is throttled due to executing PHIs, its co-located \agycr{hardware thread} running on the same physical core\agycr{,} is \emph{also throttled}. We experimentally conclude that this side-effect is due to the core's throttling mechanism halting the execution of all SMT threads for \emph{three-quarters} of the throttling period. The throttling period length is proportional to the computational intensity of the PHI any of the SMT threads executes. 

\noindent \textbf{\crj{Observation 3:} Multi-Throttling-Cores.} 
We find that when two cores execute PHIs at similar times, the throttling periods are exacerbated proportionally to the computational \crj{intensity} of each PHI across the two cores. We specifically find that this exacerbation of throttling periods occurs when two cores execute PHIs within a few hundred cycles of each other. This side-effect is because the processor power management unit (PMU) waits until the voltage transition of one core completes before starting the voltage transition of the next core. Therefore, one core can infer the computational intensity of instructions \crj{that are} being executed on another core.

Based on our \agycr{three} \crj{new} observations on current management \agy{mechanisms} in modern processors, we develop a set of covert channels, which we call \emph{\tech}.
{\tech} consists of covert channels between various execution contexts that communicate via controllable and observable throttling periods. These covert channels vary slightly depending on where the two \crj{covertly-}communicating software contexts are executing (same thread, same core, or different cores). In particular, the three covert \crj{channels} work as follows:
1) {\tech} exploits the Multi-Throttling-Thread side-effects to establish a covert channel between two execution contexts \emph{within the same hardware thread}. 
\juan{We} call this covert channel \textbf{\ta}. 
2) {\tech} exploits the Multi-Throttling-SMT side-effects to create a covert channel between two execution contexts running \ja{on} the same core but \emph{within two different SMT \ja{threads}}. We call this covert channel \textbf{\tb}. 
3) {\tech} exploits Multi-Throttling-Cores to create a covert channel between two execution contexts running \ja{on} \emph{two different physical cores}. We call this covert channel \textbf{\tc}. 
We demonstrate \tech on real modern Intel processors and find that \tech provide\agycr{s} $3Kbps$ \agycr{of covert channel capacity\crj{,} which is} more than $24\times$ the \agycr{capacity of} state-of-the-art power management-based covert channels~\cite{alagappan2017dfs, kalmbach2020turbocc, khatamifard2019powert} \agycr{and $2\times$ the capacity of}
state-of-the-art \agycr{PHI-latency-variation-based} covert channels~\cite{schwarz2019netspectre}.

\jkold{
\agytwo{B}ased on our deep understanding of the 
architectural techniques used \crj{in} current management \crj{mechanisms} to throttle the system when executing PHIs, we} propose three practical mitigation mechanisms to protect a system against known covert channels \crj{caused} by current management \agy{mechanisms} implemented in modern processors.
This work makes the following  \textbf{contributions}: 
\begin{itemize}[]

\item \jkold{We rigorously characterize real systems to develop thorough and accurate explanations for variable execution \crj{times} and frequency changes \crj{that happen} when running PHIs on modern \jr{Intel} processors \jr{(e.g., Intel Haswell \crj{\cite{6_kanter2013haswell}}, Coffee Lake \crj{\cite{coffeelake_2020}}, and Cannon Lake \crj{\cite{i38121u_cannonlake}}).}}

\item We present {\tech}, a \crj{new} set of covert channels that exploits multi-level throttling mechanisms used by the current management \agy{mechanisms} in modern processors. \jkold{These covert channels can be established between two execution contexts \crj{running} 1) on the same hardware thread, 2) \crj{on} \ja{Simultaneous Multi-Threading} (SMT) threads, and 3) \crj{across} different physical cores.} 



\item We demonstrate that  {\tech} increase\agycr{s} the channel capacity of state-of-the-art covert channels that \agycr{exploit} the variable latency of PHIs by $2\times$ \agycr{and covert channels that leverage power management mechanisms by \cj{more than} $24\times$}. 

\item We propose a \agytwo{set of} practical mitigation \agytwo{mechanisms} to protect a system against known covert channels \crj{caused} by current management mechanisms.


\end{itemize}



%% file: _02_background.tex
\section{Background}\label{sec:background}

We provide a brief background \jkt{into modern processor architectures}\footnote{We present \crj{the} \jr{Intel} client processor architecture  (e.g., Skylake \cite{anati2016inside,haj2016fine}, \ja{Kaby Lake} \cite{kabylake_2020},  \ja{Coffee Lake} \cite{coffeelake_2020}, \ja{Cannon Lake} \cite{i38121u_cannonlake}) \crj{to simplify discussions and descriptions}. Sections \ref{sec:otherprocessor} and \ref{sec:mitigations} discuss the applicability of our work to Intel server \crj{processors} and other processors.} \jkt{and their power delivery networks and electrical design limits.}



\begin{figure}[ht]
\begin{center}
\includegraphics[trim=0.5cm 0.6cm 0.5cm 0.6cm, clip=true,width=0.9\linewidth,keepaspectratio]{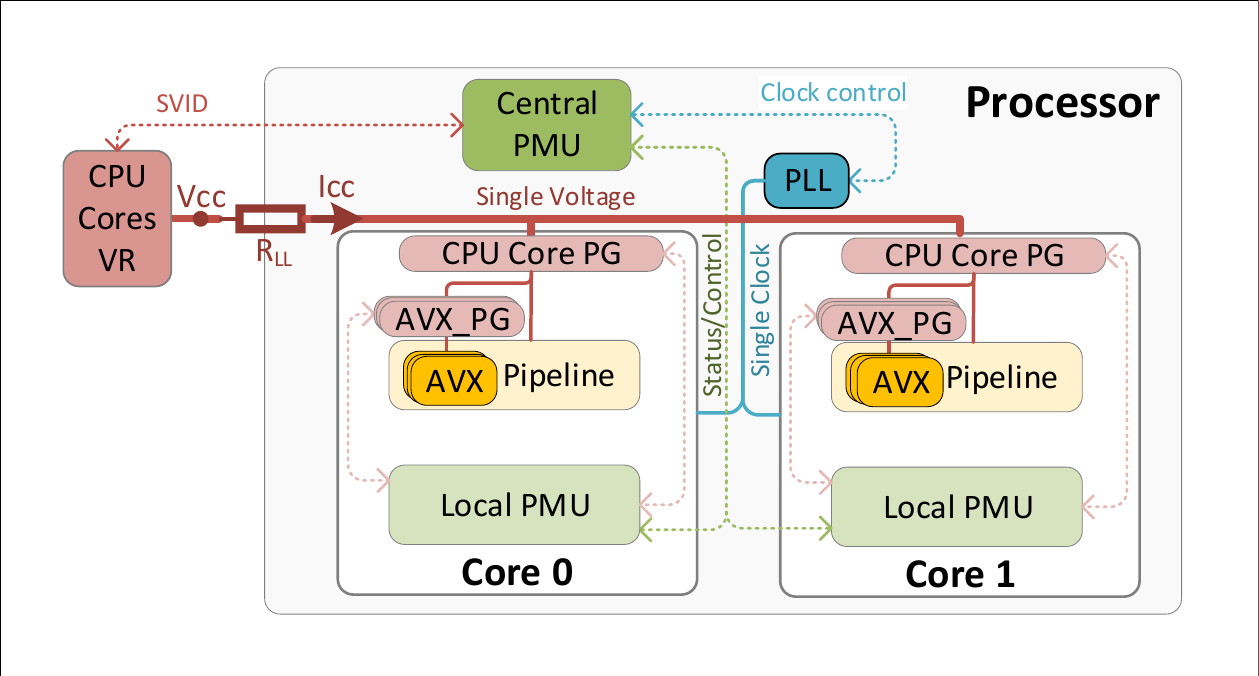}
\caption{Architecture overview of recent Intel client processors \sr{with} \crj{a focus} on CPU cores. All cores share the same voltage \crj{regulator (VR)} and clock domain. \crj{The \ja{central} power management unit (PMU) \agyfour{controls} 1) the VR using an off-chip serial voltage identification (SVID) interface, and 2) the clock phase-locked loop (PLL) using an on-chip interface.} Each \sr{CPU} core has a \ja{power-gate} (PG) for the entire core. \crj{Each} AVX unit (e.g., AVX512~\ja{\cite{mandelblat2015technology,intel_avx512}}) has a \ja{separate} PG.}\label{microarch}
\end{center}
\end{figure}

\noindent \textbf{Client \crj{Processor} Architecture.}
A high-end client processor is a system-on-chip (SoC) that typically integrates three main domains 
into a single chip: 1) compute (e.g., CPU
cores and graphics engines), 2) IO, and 3) memory \sr{system}. 
Figure \ref{microarch} show\sr{s} the architecture used in recent Intel processors (such as  Coffee Lake \cite{coffeelake_2020} and Cannon Lake \cite{i38121u_cannonlake}\sr{)}  with a focus on CPU cores. 

\noindent \textbf{Power Delivery \lois{Network} (PDN).} 
There are three commonly-used PDNs in recent high-end client processors: motherboard voltage regulators (MBVR \cite{rotem2011power,10_jahagirdar2012power,11_fayneh20164,12_howse2015tick}), integrated voltage regulators (IVR \cite{singh20173,singh2018zen,burd2019zeppelin,beck2018zeppelin,toprak20145,sinkar2013low}), and low dropout voltage regulators (LDO \cite{2_burton2014fivr,5_nalamalpu2015broadwell,tam2018skylake,icelake2020}). We present the MBVR PDN \crj{here.} Section \ref{sec:mitigations} discusses the other two PDNs.
As shown in Figure \ref{microarch}, MBVR PDN of a high-end client processor includes 1) \lois{one motherboard voltage regulator \lois{(VR)} for all CPU cores}, 2) \lois{a} \emph{load-line} resistance ($R_{LL}$\sr{)}, and  3) \ja{power-gates} for \lois{each} entire core and \crj{separately} \lois{for} each AVX unit  (e.g., AVX512~\ja{\cite{mandelblat2015technology,intel_avx512}}) inside a CPU core. All CPU cores share the same VR \cite{rotem2011power,10_jahagirdar2012power,11_fayneh20164,12_howse2015tick,haj2018energy}\sr{.} \crj{Our earlier work \cite{haj2020flexwatts} provides more \cj{background} on \cj{state-of-the-art} PDNs.}

\noindent \textbf{Clocking.} A phase-locked loop (PLL) suppl\sr{ies} the clock \sr{signal to \lois{all} CPU \lois{cores.}  \crj{All}}  CPU cores \crj{have} the same \crj{maximum} clock frequency\footnote{Intel client processors, including Haswell and \ja{Ice Lake} \sr{processor\cj{s}\cj{, which}} use \crj{fully-integrated voltage regulator (FIVR)} power delivery, have the same clock frequency domain for all cores \cite{6_kanter2013haswell,icelake2020}. } \cite{8th_9th_gen_intel,icelake2020,perf_limit_reasons,haj2018energy}.

\noindent \textbf{Power Management.} The processor includes one central power management unit (PMU) and one local PMU per CPU core. The central PMU is responsible for several power-management activities, such as dynamic voltage and frequency scaling~\ja{\cite{gonzalez1996energy,haj2018power,hajsysscale,rotem2011power}}. 
The central PMU has several interfaces with on-chip and off-chip components\sr{,} such as 
1) the motherboard VR, called serial voltage identification (\emph{SVID}) \cite{rotem2012power,rotem2015intel,haj2018power,gough2015cpu}, to control the voltage level of the VR, 
2) the PLL to control the clock frequency,
and 3) each core's local PMU for communicating power management commands and status reporting. 
The local PMU inside the CPU core is responsible for core-specific power management, such as clock gating, power gating control, thermal reporting.

\begin{figure}[ht]
\begin{center}
\includegraphics[trim=0.8cm 0.99cm 0.8cm 0.8cm, clip=true,width=1\linewidth,keepaspectratio]{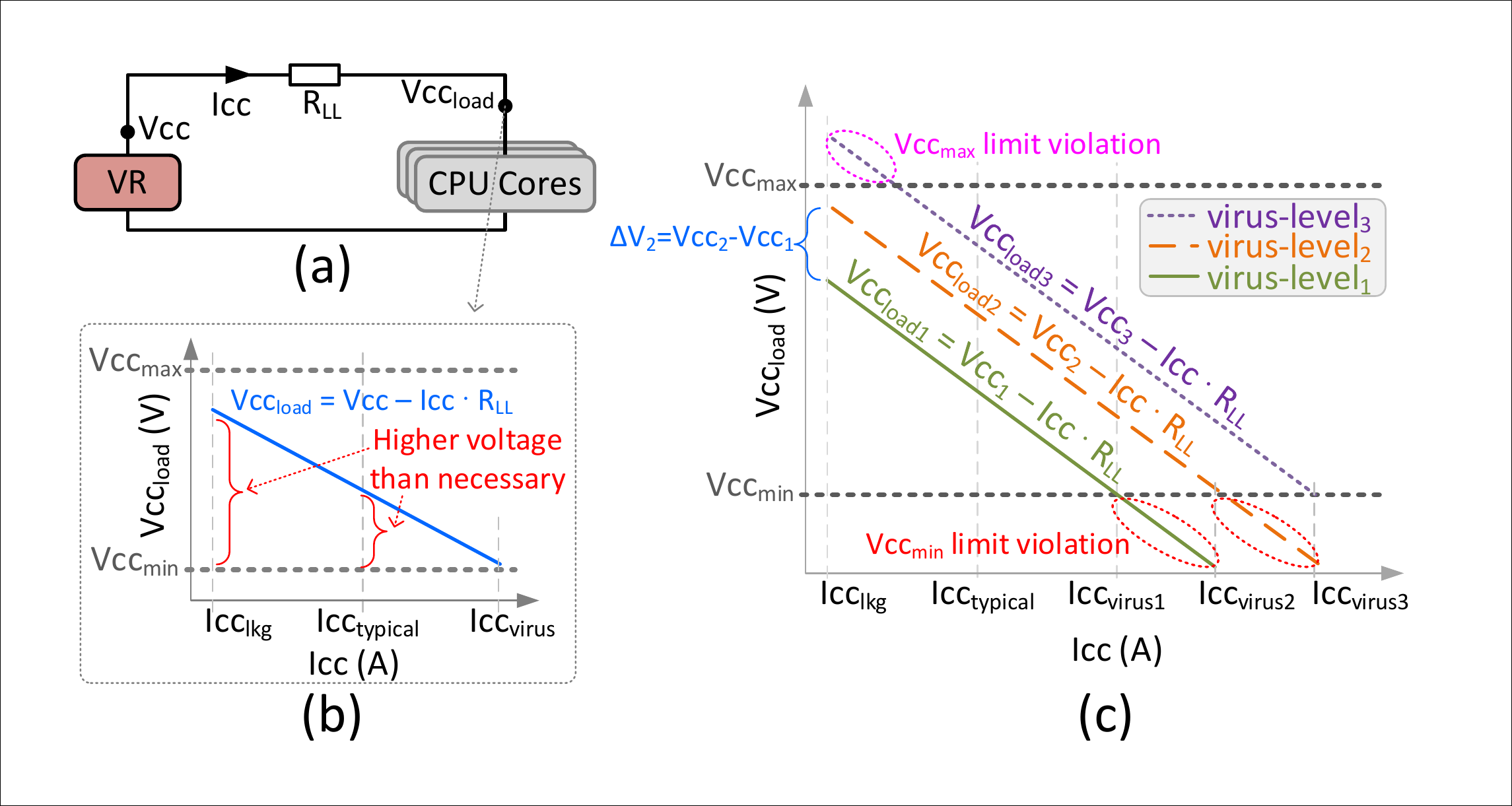}\\
\caption{Adaptive voltage guardband on modern processors. (a) Simplified Power Delivery Network (PDN) model with a load-line. \crj{(b) Voltage at the load is defined as: $Vcc_{load} =  Vcc - R_{LL}\cdot Icc$, where $Vcc$ and $Icc$ are the voltage and current at the VR output, respectively.} \crj{(c)} Multi-level load-line with three power\sr{-}virus levels. The voltage guardband is adjusted based on the power\crj{-}virus level corresponding to \sr{the} architectural state of the processor (e.g., number of active cores \crj{and instructions' computational intensity}).}\label{loadline}
\end{center}
\end{figure}

\noindent \textbf{Load-line.} \crj{Load-line} or \emph{\crj{adaptive voltage positioning}} \cite{14_module2009and,intel_avp_2009,sun2006novel,tsai2015switching} is a model that describes the voltage and current relationship\footnote{In this model, short current bursts \sr{result} in voltage droops~\ja{\cite{cho2016postsilicon,reddi2009voltage,reddi2010voltage}}\sr{,} which are filtered out by the decoupling capacitors~\cite{peterchev2006load}, while long current bursts are detected by the motherboard VR.} under a given system impedance, denoted by $R_{LL}$. \lois{Figure \ref{loadline}(a) describes a simplified power delivery network (PDN) model with a voltage regulator (VR), load-line ($R_{LL}$), and load (CPU Cores).} 
\lois{$R_{LL}$ is} typically \ja{$1.6m$--$2.4m\Omega$} for recent client processors \cite{8th_9th_gen_intel}. The voltage at the load is defined as: $Vcc_{load} =  Vcc - R_{LL}\cdot Icc$, where $Vcc$ and $Icc$ are the voltage and current at the VR output, respectively\crj{, as shown in Figure \ref{loadline}(b)}. From this equation, we can \lois{observe} that the voltage at the load input ($Vcc_{load}$) decreases when the load's current ($Icc$) increases.
Due to this phenomenon, the PMU increases the input voltage ($Vcc$), i.e., \crj{applies} a \emph{voltage guardband}, to a level that keeps the voltage at the load ($Vcc_{load}$) above the minimum functional voltage (i.e., $Vcc_{min}$) under even the most intensive load (i.e., when all \crj{active} cores \crj{are running} a workload that exercises the highest possible dynamic capacitance ($C_{dyn}$)\crj{\cj{)}. This workload is} known as a \emph{power-virus} \cite{haj2015compiler,fetzer2015managing,haj2020flexwatts} \ja{and} results in the maximum possible current ($Icc_{virus}$). A typical application consumes a lower current $Icc_{typical}$ than $Icc_{virus}$. The minimal current that the processor can consume is the leakage current ($Icc_{lkg}$) once the clocks are gated (while the supply \crj{voltage} is \emph{not} \ja{power-gated}). In all cases where the current is lower than $Icc_{virus}$, the voltage drop (i.e., $Icc \cdot R_{LL}$) is smaller than when running a power-virus, which results in a higher load voltage $Vcc_{load}$ than necessary (as shown in \crj{Figure \ref{loadline}(b)})\ja{, leading to a power loss that increases quadratically with the voltage level.}

\noindent \textbf{Adaptive Voltage Guardband, Icc$_{max}$, and Vcc$_{max}$.} To reduce the power loss \ja{resulting} from a high voltage guardband when \emph{not} running a power-virus, due to the load-line effect, modern processors define \emph{multiple levels} of power-viruses depending on the maximum dynamic capacitance ($C_{dyn}$) that an architectural state (e.g., number \sr{of} active cores, the computational intensity of running instructions) can draw. 
For each power\sr{-}virus level, the processor applies \crj{a} \emph{different} voltage guardband.  
Figure \ref{loadline}\crj{(c)} illustrates the load-line model behavior of a processor with three power-virus levels denoted by $VirusLevel_1$, $VirusLevel_2$, and  $VirusLevel_3$ (where $VirusLevel_1 < VirusLevel_2 < VirusLevel_3$). The three power-virus levels represent multiple scenarios. For example, $VirusLevel_1$, $VirusLevel_2$,  and $VirusLevel_3$ can represent one, two, or four active cores, respectively, for a processor with four cores. 
When the processor moves from one power-virus level to a higher/lower level, the processor increases/decreases the voltage by a voltage guardband ($\Delta V$). For example, when moving from $VirusLevel_1$ to  $VirusLevel_2$, the processor increases the voltage by $\Delta V_2$ as shown (\crj{in} blue \crj{text}) in Figure \ref{loadline}\crj{(c)}.

\noindent \agytwo{\textbf{Voltage and Current Limit Protection.}}
\jh{When dynamically \crj{increasing} the voltage guardband due to moving to a \crj{higher} power-virus level, the processor \crj{may} reduce the cores' frequency 1) \agytwo{to} keep the voltage within the maximum operational voltage (i.e., $Vcc_{max}$ \ja{in} Figure \ref{loadline}\crj{(c)}\ja{, pink dashed ellipse)}, and 2) \agytwo{to} keep the current ($Icc$ in Figure \ref{loadline}\crj{(b)}) consumed from the VR within the maximum current \ja{limit} ($Icc_{max}$\footnote{\jh{$Icc_{max}$ limit is the maximum current the VR must be electrically designed to support. Exceeding the $Icc_{max}$ limit can result in irreversible damage to the VR or the processor chip \cite{gough2015cpu,intel_opt_ref_manual2020}.}})~\cite{gough2015cpu,2_burton2014fivr,haj2018power,skylakex,intel_avp_2009,naffziger2016integrated,wright2006characterization,piguet2005low,zhang2014architecture}.}

\noindent \textbf{Voltage Emergency (di/dt) \jh{Avoidance}.}
\jh{When dynamically moving to a higher power-virus level, the processor increases the cores\crj{'} voltage guardband to \crj{prevent the} voltage \crj{from} dropping below the minimum \crj{operating} voltage (i.e., $Vcc_{min}$ \ja{in} Figure \ref{loadline}\crj{(c)}\ja{, red dashed \jc{ellipses}}) due to high \emph{di/dt} voltage fluctuations~\ja{\cite{joseph2003control,gough2015cpu,haj2018power,reddi2010voltage,reddi2009voltage}}\crj{.}}

\noindent \textbf{Parameters Affecting \crj{the} Voltage Guardband.}
\crj{When moving from \cj{the} power-virus level,  $VirusLevel_1$ to a higher level,  $VirusLevel_2$, the voltage guardband \jkv{($\Delta V$)}, that should be added to the supply voltage level corresponding to $VirusLevel_1$ \jkv{(\crj{$Vcc_1$})}, is correlated \crj{with} the difference in current consumption between the two levels, \jkv{($Icc_2 - Icc_1$)}. As shown in Equation \ref{eqn:guardband}, $\Delta V $ is proportional to 1) the CPU core frequency, $F$, 2) the \crj{supply voltage level corresponding to $VirusLevel_1$}, \crj{$Vcc_1$}, 3) the load-line impedance, $R_{LL}$, and 4) the difference between core dynamic \cj{capacitances} of the two power-virus \jkv{levels}, $C_{dyn2} -  C_{dyn1}$}.    

\begin{equation}\label{eqn:guardband}
\resizebox{.7\hsize}{!}{$
\begin{array}{c}
\Delta V = Vcc_2 - Vcc_1 \approx (Icc_2 - Icc_1) \cdot R_{LL} \\[\jot]
= (C_{dyn2} \cdot Vcc_1 \cdot F -  C_{dyn1} \cdot Vcc_1 \cdot F) \cdot R_{LL} \\[\jot]
= (C_{dyn2} -  C_{dyn1}) \cdot Vcc_1 \cdot F \cdot R_{LL}
\end{array}
$}\end{equation}

The dynamic capacitance of the CPU cores at a given point in time
is correlated \crj{with} the 1) number of \emph{active cores}, 2) the \emph{capacitance}, $C$, of each core, and 3) the node \emph{activity factor} of each core. The capacitance, C, of a core is a function of the computational intensity (e.g., MOV, OR, ADD, MULL, FMA) and width (e.g., 64\crj{-}bit scalar, 128\crj{-}bit vector, 256\crj{-}bit vector, 512\crj{-}bit vector) \lois{of the instructions executed by} the cores.   

\noindent \textbf{Power Gating.}
Power gating is a circuit-level technique to eliminate leakage power of an idle circuit \cite{hu2004microarchitectural,haj2018power,gough2015cpu}. 
Typically, the wake-up latency from the \ja{power-gated} state can take a few cycles to tens of cycles \cite{kahng2013many,gough2015cpu}. However, to reduce the worst-case peak current and voltage noise on the power delivery (e.g., di/dt noise \crj{\cite{larsson1997di,gough2015cpu,haj2018power}}) when waking up a power-gate, \lois{the \ja{power-gate} controller applies} a \emph{staggered} wake up  \crj{technique} \cite{agarwal2006power} \lois{that} takes tens of nanoseconds (typically, $10$--$20ns$) \cite{kahng2013many,akl2009effective,kahng2012tap}.

%% file: _03_limitations.tex
\section{Limitations of Prior Work}
\label{sec:limitations}
\agytwo{\crj{We describe} two state-of-the-art covert channels}, NetSpectre~\cite{schwarz2019netspectre} and TurboCC~\cite{kalmbach2020turbocc}\sr{, and their major limitations based on our rigorous characterization} (\ja{which is provided in} Section \ref{sec:profiling}).


Schwarz \emph{et al.} propose \agytwo{NetSpectre~\cite{schwarz2019netspectre}, a mechanism that builds a covert channel by exploiting the variation \crj{in} AVX2~\cite{hammarlund2014haswell} instructions' execution times \crj{due to throttling mechanisms}.}
\agytwo{\crj{We} \sr{observe} that NetSpectre \sr{has} three \lois{main} limitations}.
\juan{First,} \crj{the} NetSpectre  covert channel can be established \sr{\emph{only}} between two execution contexts on the \emph{same hardware thread}, whereas \agytwo{our rigorous characterization demonstrates} that the throttling side-effect can be observed across \sr{both} threads of different cores (Section \ref{sec:crosscore_multi_level}) \sr{and \ja{Simultaneous Multi-Threading} (SMT) threads (Section \ref{sec:crossthread_multi_level})}. Therefore, NetSpectre has a  \ja{narrower} and \ja{more} limited attack vector compared to our work.  
\juan{Second, NetSpectre} uses a single-level throttling side-effect (i.e., whether or not the thread is throttled) to communicate confidential data. \lois{Our rigorous characterization shows} that the throttling side-effect can result in multi-level (up to five) throttling periods \crj{based on} the power requirements of the PHIs \crj{that are} being executed (\lois{see} \jh{Sections \ref{sec:crosscore_multi_level} and  \ref{sec:crossthread_multi_level}})\crj{. \tech  utilizes the multi-level throttling periods to transmit two bits \jkv{per} communication transaction while NetSpectre sends only a single bit per transaction}. 
\jh{Therefore, NetSpectre can use \sr{only} \emph{half} of the \crj{covert channel} bandwidth \sr{provided by} \ja{the} \crj{throttling side-effect of current management mechanisms}.}
\juan{Third, NetSpectre} does \emph{not} identify the true source of throttling. \crj{As a result,} NetSpectre \jr{does not} propose \sr{any} mitigation techniques. 
\sr{NetSpectre} \je{hypothesizes that} \agytwo{the source of throttling} \crj{is} the power-gating of the AVX2 execution unit, while we \agytwo{demonstrate}, in Section \ref{sec:not_power_gating}, that ${\sim}99\%$ of the \crj{throttling period} is due to voltage transitions initiated by current management \agy{mechanisms}\crj{. Thus,} we propose, in Section \ref{sec:mitigations}, \crj{practical and effective mitigation techniques}.

Kalmbach \emph{et al.}~\cite{kalmbach2020turbocc} propose TurboCC,  a mechanism that \agytwo{creates cross-core covert channels by exploiting the core frequency throttling when executing PHIs (e.g., AVX2) \juan{at Turbo frequencies~\cite{rotem2012power,rotem2013power}}. 
\lois{\sr{We} identify two main limitations in} TurboCC.} 
\juan{First,} TurboCC focuses \emph{only} on the slow side-effect of frequency throttling \crj{that happens} when \juan{executing} PHIs at \emph{Turbo} frequencies, which takes tens of milliseconds. \lois{In our work, we} observe and utilize a \ja{cross-SMT-thread} and \ja{cross-core} immediate side-effect (\ja{that occurs in} tens of microseconds, \juan{i.e.,} three orders of magnitude faster than \jc{the} frequency changes TurboCC \ja{relies on}) that \ja{happens} at \emph{any} frequency (not only Turbo \crj{frequencies}), \jh{as we discuss in Sections \ref{sec:crosscore_multi_level} and  \ref{sec:crossthread_multi_level}}. 
\juan{Second, TurboCC does} not uncover the real reason behind the vulnerability. The authors \ja{hypothesize} that the side-effect of frequency throttling is due to thermal management mechanisms \ja{\cite{rotem2013power, singla2015predictive, rotem2011power, rotem2015intel,brooks2001dynamic}}, while we observe that \sr{this} side-effect is due to current management \agy{mechanisms} \crj{\cite{gough2015cpu,2_burton2014fivr,haj2018power,skylakex,intel_avp_2009,naffziger2016integrated,wright2006characterization,piguet2005low,zhang2014architecture}} that also \crj{exist} in thermally\crj{-}unconstrained systems \crj{or when thermals are not a problem during execution in a system} (Section \ref{sec:iccmax_vccmax}). 

To our knowledge, no previous work  
1) provides \crj{and analyzes the fundamental} reasons \juan{why} each \crj{type of} throttling occurs when executing PHIs \crj{(Section \ref{sec:profiling})}; 
2) comprehensively exploits the multi-level throttling side-effects of the current management \agy{mechanisms} \cj{within a thread, across SMT, and across cores} \crj{(Section \ref{sec:covert_overview})}; 
\lois{and 3) proposes practical and effective mitigation techniques \crj{(Section \ref{sec:mitigations})}.}

%% file: _04_technique.tex
\section{\tech Overview}\label{sec:covert_overview}
\noindent \textbf{Attacker Model.} We assume a standard threat model for a covert channel attack~\crj{\cite{kocher2019spectre, khatamifard2019powert, alagappan2017dfs, kalmbach2020turbocc, schwarz2019netspectre,millen1987covert,wu2014whispers,wang2006covert}}.
\sr{Such a threat model consists} of two malicious user-level \crj{attacker} applications, sender and receiver, that \emph{cannot} communicate legitimately through overt channels. 
The sender has legitimate access to sensitive information (such as a secret key). However, the sender does not have access to any overt channel for data communication (e.g., system calls or inter-process communication). On the other hand, the receiver has access to an overt channel with the attacker, but does not have access to sensitive information. The \crj{common} goal of the sender and the receiver is to exfiltrate sensitive data despite the lack of an overt \ja{communication} channel between them. 

\noindent \textbf{Attack.} Based on our characterization \crj{of the throttling side-effects of the current management
mechanisms (Section \ref{sec:profiling})}, we build three high-throughput covert channels. These covert channels exploit \emph{three} side-effects \jh{(on the same hardware thread, across SMT threads, and across cores)} of \jh{core execution} throttling mechanisms on modern Intel processors when executing power-hungry instructions (PHIs) \jh{at any core  frequency}. 
We first briefly discuss these key side-effects 
in the context of \tech covert channels and later present our full \jk{real} \crj{system} characterization results in Section~\ref{sec:profiling}.

\begin{figure}[ht]
\begin{center}
\includegraphics[trim=0.6cm 0.6cm 0.6cm 0.8cm, clip=true,width=0.9\linewidth,keepaspectratio]{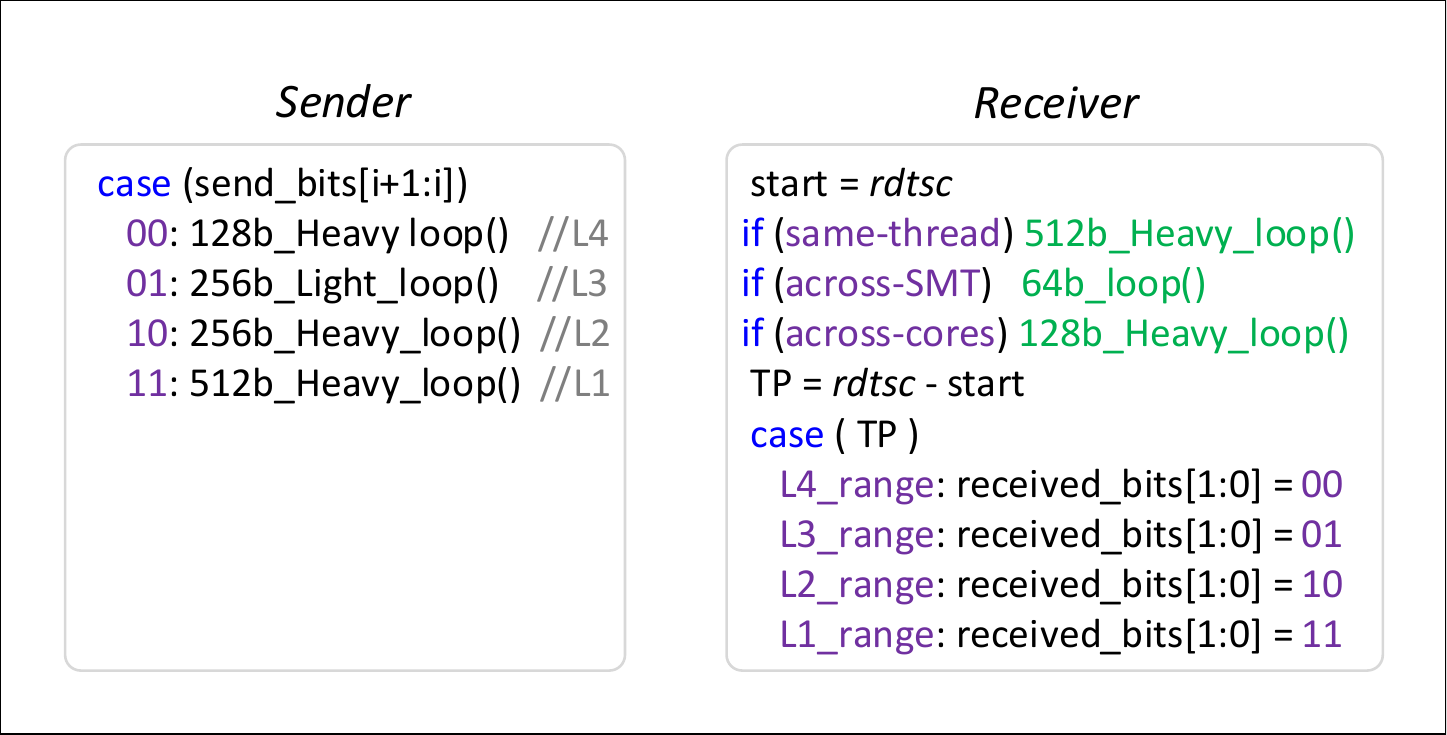}
\caption{\ja{Common pseudo-code of two} execution contexts (\texttt{Sender} and \texttt{Receiver}) \ja{that can be} located on the same hardware \jk{thread, across SMT threads, and across cores,} communicating via \ta, \tb, and \tc, respectively.}\label{coverts}
\end{center}
\end{figure}

\crj{Figure \ref{coverts} demonstrates \ja{the} \emph{common pseudo-code} of the three covert channels. \texttt{Sender} code can send two bits of a secret (\texttt{send\_bits[i+1:i]}) per communication transaction. The  \texttt{Sender} executes a PHI in a loop (e.g., \cj{a} few thousand loop iterations). The PHI's computational intensity level depends on the value of the two secret bits. In particular, there are four computational intensity levels (L1--L4) correspond\cj{ing} to four instruction types, i.e., \ja{\texttt{128b\_Heavy}}, \texttt{256b\_Light}, \texttt{256b\_Heavy}, and \ja{\texttt{512b\_Heavy}}. We define \emph{Heavy} instructions to include any instruction that requires the floating-point unit (e.g., ADDPD, SUBPS) or any multiplication instruction, while \emph{light} instructions include all other instructions (e.g., non-multiplication integer arithmetic, \jc{logic}, shuffle and blend instructions).} 

\crj{The \texttt{Receiver} can then read the secret data by simply executing one of three instruction types in a loop (e.g., \cj{a} few thousand loop iterations) while measuring its \ja{own} \emph{throttling period} (\texttt{TP}) using the \texttt{rdtsc} instruction.   Depending on the \texttt{Sender} and \texttt{Receiver} locations (\jkv{e.g.}, same  hardware  thread,   across  SMT  threads, or across cores) the \texttt{Receiver} \cj{executes one} of three instruction types  (i.e., \ja{\texttt{512b\_Heavy}}, \texttt{64b}, or \ja{\texttt{128b\_Heavy}}). The \texttt{Receiver} decodes the transmitted bits by the \texttt{Sender} based on the \texttt{TP} range \cj{it measures}. \ja{Since the \texttt{TP} range is predictably dependent on the instruction executed by the \texttt{Sender}, the \texttt{Receiver} can infer the instruction executed by the  \texttt{Sender}.}}

\begin{figure}[!ht]
\begin{center}
 \includegraphics[trim=0.5cm 0.7cm 0.5cm 0.7cm, clip=true,width=1\linewidth,keepaspectratio]{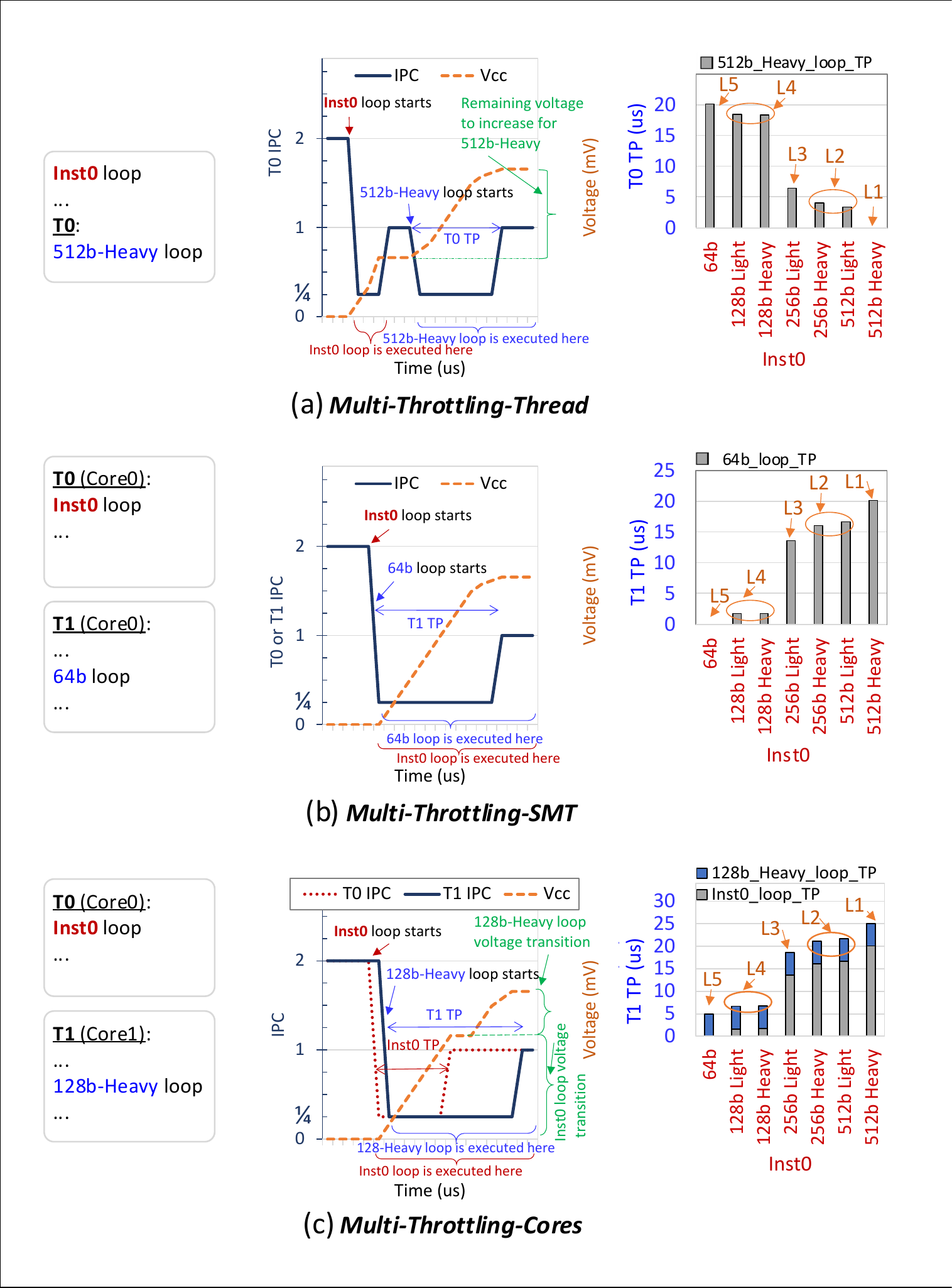}
\caption{\crj{Illustrating the three throttling side-effects: (a) In Multi-Throttling-Thread, the \texttt{Inst0} loop \cj{increases} the voltage based on \ja{the} power-level \ja{required to execute \texttt{Inst0}} \cj{and the} subsequent \texttt{512b\_Heavy} loop, which requires the worst-case voltage, \ja{increases} the voltage \ja{further beyond} the level reached by \texttt{Inst0} loop\cj{.} \ja{Thus, the throttling period (\texttt{TP}) is \jc{dependent} on the type of \texttt{Inst0}.} (b) In Multi-Throttling-SMT, when executing the \texttt{Inst0} loop on thread \texttt{T0}, the IPC of both SMT threads is \jc{reduced by $75\%$}, therefore, the \texttt{TP} depends on the \ja{type of} \texttt{Inst0}. (c) In Multi-Throttling-Cores,  running \cj{the} \texttt{Inst0} loop on \texttt{Core0} results in throttling the core and requesting a voltage increase. Subsequent execution of \cj{the} \texttt{128b\_Heavy} instruction on \texttt{Core1} results in throttling the core and \cj{requesting a} voltage increase. Since the voltage regulator (VR) is busy with \texttt{Core0} voltage transition,  \texttt{Core1} remains throttled until  both cores complete \cj{their} voltage transitions.}}
\label{throttling_effects}
\end{center}
\end{figure}

\vspace{25pt}
\subsection{Covert Channel 1: IccThreadCovert}
\sr{The first covert channel is \ta, which exploits the Multi-Throttling-Thread side-effect to establish a covert channel}.

\subsubsection{Side-Effect 1: Multi-Throttling-Thread}
We find that executing \sr{PHIs} results in a multi-level (i.e., one of several discretized) throttling period lengths proportional to the computational intensity of the instructions. \jk{Observing the variation in instructions per cycle (IPC) during execution allows \jd{the} \jc{\texttt{Receiver}} to determine the throttling period. Based on this,} \jd{the} \jc{\texttt{Receiver}} can determine the computational intensity of recently executed instructions \jc{by \jd{the} \texttt{Sender}}.
We find that \ja{the} throttling period is primarily due to \ja{and \jc{dependent} on} the higher voltage guardband requirement of instructions with higher computational intensity, which requires more time to ramp up the voltage level of the voltage regulator. For example, executing a 512-bit vector multiplication instruction results in a longer throttling period than executing a 256-bit vector bitwise OR instruction. 

\ja{We find that} executing an instruction with high computational intensity results in a throttling period proportional to the difference in voltage requirements of 1) the currently executing instruction and 2) the previously executed instruction. 
\crj{Figure~\ref{throttling_effects}(a) illustrates this. 
At the \cj{beginning,} the processor \cj{executes} scalar instructions with $IPC=2$. Later \texttt{Inst0} loop \jkv{starts} executing (we assume IPC of \texttt{Inst0} is $1$). \texttt{Inst0} loop is throttled ($IPC=\sfrac{1}{4}$) while ramping the voltage ($Vcc$) to accommodate \texttt{Inst0}\cj{'s} computational intensity. Once the target voltage is reached, the throttling is stopped ($IPC=1$).
When \texttt{512b\_Heavy} (we assume $IPC=1$) loop is executed, \jkv{it is} first throttled ($IPC=\sfrac{1}{4}$) while ramping the voltage ($Vcc$) to accommodate \texttt{512b\_Heavy}. The throttling period (\texttt{T0} \texttt{TP} in Figure~\ref{throttling_effects}(a)) before executing the \texttt{512b\_Heavy} instructions with full IPC (i.e., $IPC=1$)  is dependent on the computational intensity of \texttt{Inst0} loop, since the remaining voltage \ja{required to execute a} \texttt{512b\_Heavy} instruction depends on the previous voltage level that was reached when \texttt{Inst0} loop was executed. The bar chart \cj{on the right part of Figure~\ref{throttling_effects}(a)} shows how the throttling period (\texttt{T0} \texttt{TP}) changes with \jk{the} computational intensity of \jkv{the} \texttt{Inst0} loop. \ja{For} instructions \ja{(\texttt{Inst0})} with lower computational intensity \ja{that} require \jk{lower} voltage \ja{level to execute}, the \ja{required} voltage \ja{increase needed to next execute} a \texttt{512b\_Heavy} loop is larger and \ja{hence the \texttt{TP} is longer (see L5 and L1 in the bar chart)}. An observer of the \jc{\texttt{TP}} of the \texttt{512b\_Heavy} loop can \jc{thus} infer the computational intensity of the previously executed instructions in \cj{the} \texttt{Inst0} loop.}


\subsubsection{IccThreadCovert}
\label{sec:IccThreadCovert}

\emph{\ta} exploits the Multi-Throttling-Thread to build a covert channel between two execution contexts, \texttt{Sender} and \texttt{Receiver}, running on the same hardware thread. 
Figure \ref{coverts} demonstrates how \texttt{Sender} code can send two bits of a secret (\texttt{send\_bits[i+1:i]}) per communication transaction. The  \texttt{Sender} executes PHIs with a computational intensity level (\sr{one out of four levels,} L1--L4) depending on the value of the two secret bits \ja{(as shown in Figure \ref{coverts})}. While the current management mechanisms \sr{adjust} the supply voltage to the core to provide enough power to execute the \jd{PHI} \jc{loop} of the \texttt{Sender}, the IPC is reduced to \crj{$\sfrac{1}{4}$\ja{th}} of its \ja{baseline} value (i.e., $IPC=1$). This IPC throttling lasts for a length \ja{of} time proportional to the computational intensity of the  PHI \jc{that the \texttt{Sender} executes}.   


The \texttt{Receiver} can then read the secret data by simply executing a \texttt{512b\_Heavy} loop \cj{(i.e., same-thread \ja{in Figure \ref{coverts}})} and measuring its throttling period (\texttt{TP}). The higher the power required by the \jc{\jd{PHI} loop}  executed by the \texttt{Sender}\jk{,} the \sr{shorter} the \texttt{TP} experienced by the \texttt{Receiver} will be, \sr{as} the voltage supplied to the core was already partially ramped up to execute the  \crj{\jd{PHI} loop} \jd{of the \texttt{Sender}}. \sr{The} \texttt{Receiver} can identify the two bits of secret sent by the \texttt{Sender} \sr{based on the length of the \texttt{TP}}. The \texttt{Sender} can initiate sending additional data after waiting a \sr{certain period of} time. 
\crj{We find that a waiting period} of approximately $650us$ is required after the last time the thread executes a PHI. We refer to this waiting \sr{period} as \emph{reset-time}. The need for a waiting period is due to the fact that the processor keeps a hysteresis counter that keeps the voltage at a high level corresponding to the highest power PHI executed within the reset-time frame. If no \sr{executed} PHIs \crj{are} within a $650us$ time frame, the processor reduces the voltage to the baseline voltage level. After the reset-time \ja{passes}, subsequent \sr{executions} of PHIs result \sr{in} the \crj{three} side-effects that we describe in this section \crj{(Section \ref{sec:covert_overview})}.

\subsection{Covert Channel 2: IccSMTcovert}
The second \crj{covert} channel is \tb, which exploits the Multi-Throttling-SMT side-effect\crj{.}

\subsubsection{Side-Effect 2: Multi-Throttling-SMT}
We find that\sr{,} in a processor with \ja{Simultaneous Multi-Threading} (SMT)\crj{\cite{tullsen1995simultaneous}}\sr{,} when a thread is throttled due to executing PHIs, the co-located (i.e., running on the same physical core \crj{but different SMT context}) hardware thread is \emph{also throttled}. Depending on the length of the throttling period, a co-located hardware thread can infer the computational intensity of instructions that have recently been executed \crj{in the throttled thread}. 
We discover that co-located hardware threads are throttled \crj{\emph{together}} because the throttling mechanism in the core pipeline blocks the front-end to back-end interface during \emph{three-quarters} of the \texttt{TP} for the \emph{entire core}\crj{, as we show in Section \ref{sec:crossthread_multi_level}}.
Figure \ref{throttling_effects}(b) demonstrates how two SMT threads, T0 and T1, are throttled by a single-core executing PHIs. T0 executes \ja{the} \texttt{Inst0} loop, while T1 executes a scalar \texttt{64b} instruction loop.
The \je{IPCs} of both T0 and T1 \je{drop} once the co-located thread, T0, starts executing \texttt{Inst0} loop. The throttling period depends on the computational intensity  of \texttt{Inst0} (executed by T0), which determines the voltage level to which the processor needs to increase the \ja{supply voltage}. The \ja{higher the} voltage \ja{level} required, the longer the throttling period required (\crj{as} shown by the bar chart \crj{that depicts our measurements for \texttt{TP}}). 

\subsubsection{IccSMTcovert}

\emph{\tb} exploits the Multi-Throttling-SMT side-effect to provide a covert channel between two SMT threads. 
Figure \ref{coverts} shows \texttt{Sender} code that \sr{sends} two bits of a secret (\texttt{send\_bits[i+1:i]}) \crj{per communication transaction}. The \texttt{Sender} executes \jd{a} \jd{PHI} \jc{loop} with a computational intensity level (L1--L4) depending on the two secret bits\crj{'} values. This code results in throttling of the entire core (frond-end to back-end interface) in which the IPC\crj{s of} both SMT threads \je{drop to} \crj{$\sfrac{1}{4}$\je{th}} of \crj{their} usual \crj{values (i.e., $IPC=1$)}. The throttling period (\texttt{TP}) is proportional to the computational intensity of the executed PHI \jc{by the \texttt{Sender}}, \crj{similarly to} as described for \ta.


The \texttt{Receiver} code in Figure \ref{coverts} on \crj{a} co-located hardware  thread \crj{(i.e., \emph{across-SMT} \ja{in Figure \ref{coverts}})}, \crj{executes} a scalar loop (denoted by \texttt{64b\_loop}) while measuring the \texttt{TP} of the loop. Based on the \texttt{TP} time, the \texttt{Receiver} determines the value of the two bits sent by the \texttt{Sender} code. The \texttt{Sender} can initiate sending additional two bits after waiting for a \emph{reset-time}, which is the same as the one used for IccThreadCovert. 

\subsection{Covert Channel 3: IccCoresCovert}
\sr{The third covert channel is \tc, which exploits the Multi-Throttling-Cores side-effect of PHIs\cj{.}}

\subsubsection{Side-Effect 3: Multi-Throttling-Cores}\label{sec:MultiThrottlingCores}
We find that when two cores execute PHIs at similar times, the throttling periods are exacerbated proportionally to the computational \ja{intensity} of each PHI \ja{executed in each core}. One core can thus infer the computational intensity of instructions being executed on another core. 
We specifically find that this exacerbation of the throttling period occurs when two cores execute PHIs within a few hundred cycles of each other. This increase \crj{in} the throttling period is because the processor\sr{'s} \ja{central} PMU waits until the voltage transition \sr{for} core A \cj{to} complete before starting the voltage transition for core B. 


Figure \ref{throttling_effects}(c) shows how the throttling period across two threads, T0, and T1, running on two different cores, \jc{Core0} and \jc{Core1}, can affect each other. When T0 initiates the execution of \texttt{Inst0} loop, and T1 shortly after initiates the execution of a \texttt{128b\_Heavy} PHI \ja{loop}, the throttling period (\texttt{TP}) of T1 depends on the computational intensity of the  instructions executed by the T0 thread. 

\subsubsection{IccCoresCovert}

\emph{\tc} exploits the Multi-Throttling-Cores side-effect of PHIs to provide a covert channel between two threads in two different physical CPU cores. 
Figure \ref{coverts} shows how a \texttt{Sender} thread can send two bits of secret information (\texttt{send\_bits[i+1:i]} per transaction with \hp{six} steps. 
\crj{1)} \hp{Synchronize  \crj{(not shown in Figure \ref{coverts})} \texttt{Sender} and \texttt{Receiver} threads (see detail below)}.
\crj{2)} The \texttt{Sender} thread executes \crj{a} \jd{PHI} \crj{loop} with a computational intensity level (L1--L4) depending on the two secret bits\crj{'} \crj{values}. When executing a PHI, the processor throttles the execution and initiates a voltage ramp-up command to the voltage regulator. 
\crj{3)} A \texttt{Receiver} thread  executes its \jd{PHI} loop \cj{(i.e., \jd{the} \ja{\texttt{128b\_Heavy\_loop}}\ja{, corresponding} to \emph{across-cores} \ja{in Figure \ref{coverts}})} code while the \texttt{Sender} code is throttled (waiting for the voltage transition to complete). 
\crj{4)} The \texttt{Receiver} continues to be throttled until the voltage transition of the \texttt{Sender} is complete. The duration of the \texttt{Sender} voltage transition depends on the computational intensity level of the PHI the \texttt{Sender} \ja{executes}. 
\crj{5)} Once the voltage transition of the \texttt{Sender} is complete, the processor starts a voltage transition corresponding to the computational intensity of the \jd{PHI} loop that \ja{the} \texttt{Receiver} \cj{executes} (i.e.,  \texttt{\ja{128b\_Heavy\_loop}}). This process effectively throttles the  \texttt{Receiver} code by a \texttt{TP} time proportional to the computational intensity of the \jc{\jd{PHI} loop} executed by the \texttt{Sender}. 
\crj{6)} The \texttt{Receiver} measures its \texttt{TP} \crj{using \ja{the} \texttt{rdtsc} instruction} \crj{to} determine the two bits sent by the \texttt{Sender}\sr{,} as shown in \crj{the \texttt{Receiver} code of} Figure \ref{coverts}.

\subsubsection{Thread Synchronization} \hp{To correctly transfer data between the \texttt{Sender} and the  \texttt{Receiver} threads, it is essential to synchronize their operations precisely. One common way to perform this synchronization is by using the wall clock~\cite{pessl2016drama}, where each thread can obtain the wall clock using \texttt{rstsc} instruction. To accurately synchronize the \texttt{Sender} and \texttt{Receiver} threads, each thread waits (e.g., executes \texttt{rdtsc} in a busy loop at the beginning of its code) for fixed points in time  before starting to execute its actual covert-channel code.}

%% file: _05_motivation.tex
\section{Throttling Characterization}\label{sec:profiling}


To understand the true architectural techniques used by a current management mechanism to throttle the system when executing PHIs, we experimentally study three modern Intel processors (i.e., Intel Haswell \cite{6_kanter2013haswell}, Coffee Lake \cite{coffeelake_2020}, and Cannon Lake \cite{i38121u_cannonlake}). 
First, we describe our experimental methodology. Second, we study the side-effects of running PHIs on supply voltage, current, and frequency. Third, we investigate whether the real cause \jk{of throttling} is power-gating as hypothesized in prior work \cite{schwarz2019netspectre}. After eliminating power-gating as a potential cause, we explore whether thread throttling occurs due to micro-architectural throttling techniques implemented inside the core and examine \jk{the} throttling side-effects. Finally, we reveal a multi-level throttling side-effect between cores due to \jkx{the} shared voltage regulator (VR) between cores.   

\subsection{Experimental Methodology}
\label{sec:method}
We characterize current management mechanisms
using real systems with modern Intel processors: Intel Haswell (Core i7-4770K \cite{6_kanter2013haswell}, four cores), Coffee Lake (Core i7-9700K \cite{coffeelake_2020}, eight cores), and Cannon Lake (Core i3-8121U \cite{i38121u_cannonlake}, two cores).

\begin{figure}[ht]
\begin{center}
\includegraphics[trim=0.6cm 0.6cm 0.6cm 0.6cm, clip=true,width=1\linewidth,keepaspectratio]{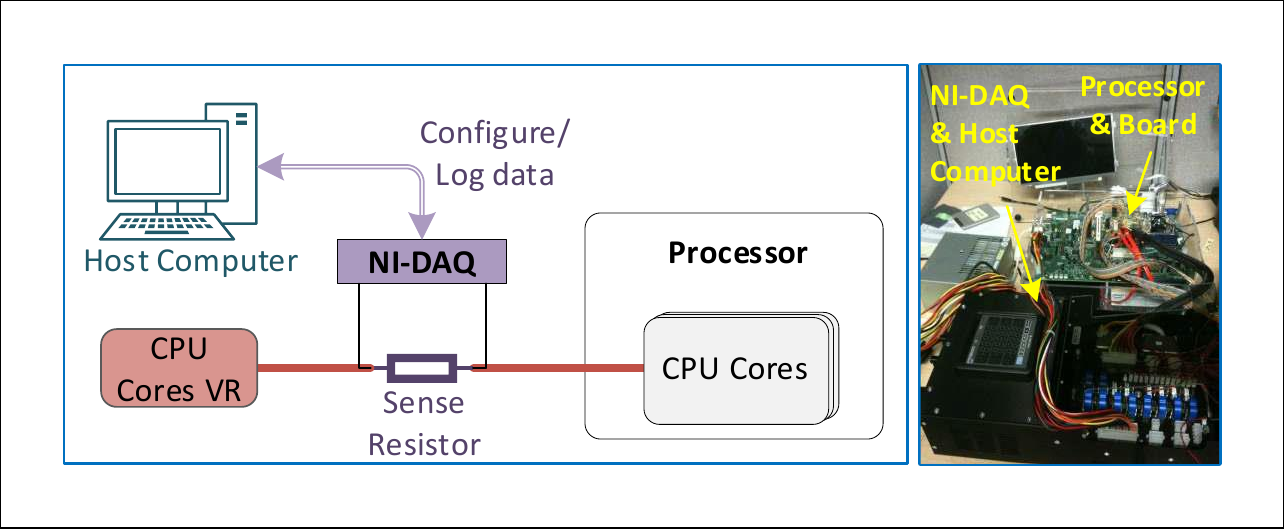}
\caption{\ja{Voltage and current measurement infrastructure using a National Instruments Data Acquisition (NI-DAQ) card. Differential cables transfer multiple measurement signals from the CPU cores' voltage regulator output wires and sense resistors on the motherboard to the NI-DAQ card in the host computer that collects the measurement data.}}\label{measurements_setup}
\end{center}
\end{figure}

\noindent \textbf{Voltage and Current Measurements.} We measure the voltage and current of the CPU core while executing PHIs with a National Instruments Data Acquisition (NI-DAQ) card (NI-PCIe-6376~\cite{NIDAQ}), whose sampling rate reaches up to 3.5 Mega-samples-per-second (MS/s). Differential cables transfer multiple measurement signals from the CPU cores' voltage regulator output wires \ja{}{and sense \jb{resistors}} on the motherboard to the NI-DAQ card in the host computer that collects the power measurements\ja{, as depicted in Figure \ref{measurements_setup}}. The power measurement accuracy of the NI-PCIe-6376 is $99.94\%$ \cite{NIDAQ}. For more detail, we refer the reader to the National Instruments manual \cite{NIDAQ}.

\noindent \textbf{Performance Counters and Micro-benchmarks.}
We customize multiple micro-benchmarks of the Agner Fog measurement library \cite{pmc_agner2020} and implement Performance Monitoring Counters (PMCs) that we track in our experiments.

\subsection{Voltage Emergency Avoidance Mechanism} 
\begin{figure}[ht]
\begin{center}
\includegraphics[trim=0.6cm 0.6cm 0.6cm 0.75cm, clip=true,width=1\linewidth,keepaspectratio]{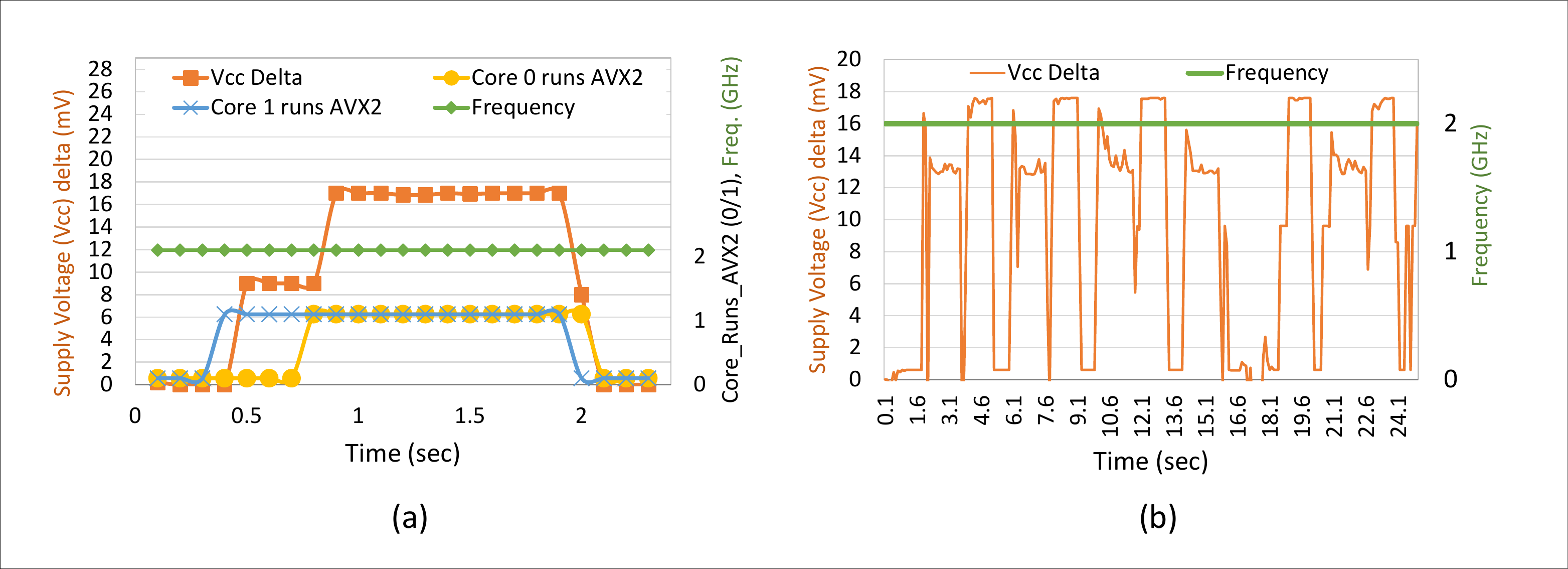}
\caption{\jk{Change in} supply voltage ($Vcc$, left y-axis) and cores' frequency (right y-axis) when (a) multiple cores execute code with AVX2 instructions (\jk{as shown by Core\_Runs\_AVX2, on the} right y-axis), and (b) executing 454.calculix from SPEC CPU2006~\cite{SPEC2018} with AVX2 on two cores. Both (a) and (b) run at 2GHz clock frequency, which is significantly lower than the baseline frequency (3.6GHz).}\label{didt}
\end{center}
\end{figure}

The voltage emergency avoidance mechanism 
prevents the voltage from dropping below the minimum operating voltage (i.e., $Vcc_{min}$) due to high \emph{di/dt} voltage fluctuations, as discussed in Section \ref{sec:background}.
We study the impact of PHI instructions on the supply voltage of the CPU cores (i.e., $Vcc$). To do so, we track the \jk{change in} $Vcc$ during \emph{two} experiments on a two-core Coffee Lake system (i7-9700K): \cj{1) two CPU cores executing \jk{code} that includes AVX2 phases, and 2) two CPU cores executing 454.calculix from SPEC CPU2006~\cite{SPEC2018}, compiled with auto-vectorization to AVX2 \cite{intel_icc_compiler}. In both experiments, the CPU core clock frequency is set to $2GHz$, which is significantly lower than the baseline frequency ($3.6GHz$)}.
Figure~\ref{didt}(a) shows the \jk{change in $Vcc$ (i.e., $Vcc$ \jkx{delta}, shown on the left y-axis), relative to the starting $Vcc$, (i.e., \ja{$788mV$}),} when multiple cores execute code with AVX2 instructions, and Figure~\ref{didt}(b) shows the \jk{change in $Vcc$} when executing 454.calculix \jk{under the same setup}.   

We make five key observations from Figure \ref{didt}(a). First, when core 1 (blue plot) begins executing AVX2 instructions ($time=\SI{0.4}{\second}$), supply voltage increases by approximately \SI{8}{\milli\volt}. Second, when core 0 (yellow plot) also begins executing AVX2 instructions ($time=\SI{0.8}{\second}$), supply voltage increases by an additional \SI{9}{\milli\volt}. Third, when core 1 stops executing AVX2 instructions ($time=\SI{2.0}{\second}$), supply voltage reduces by $8mV$. Fourth, when core 0 also stops executing AVX2 instructions ($time=\SI{2.1}{\second}$), supply voltage returns to its original value \jk{(i.e., \ja{$788 mV$})}. Fifth, the clock frequency of the cores (green plot) remains the same throughout the entire execution of \jk{code on both cores}. This is also reflected in Figure \ref{didt}(b), where throughout the execution of the 454.calculix workload, the processor only adjusts the supply voltage depending on the code phases (Non-AVX vs. AVX2) of each of the two cores. 

We conclude that 1) the processor adjusts the supply voltage proportionally to the number of cores executing PHIs (e.g., AVX2) to prevent voltage emergencies (i.e., di/dt noise) due to the instantaneous high current that PHIs consume (as discussed in Section \ref{sec:background}), and 2) core frequency is \emph{not} affected when the processor runs at a low clock frequency relative to its baseline frequency, since such a frequency does not violate the processor power delivery current and voltage limits.  

\noindent \textbf{Key Conclusion 1.}
\textit{A modern processor employs the voltage emergency (i.e., di/dt noise) avoidance mechanism, a current management mechanism that prevents the core voltage from dropping below the minimum operational voltage limit (\ja{$Vcc_{min}$}) when executing PHIs.}


\subsection{Maximum Icc/Vcc Limit Protection Mechanisms}\label{sec:iccmax_vccmax} 
In order to demonstrate the impact of executing PHIs (e.g., AVX2) on supply current (i.e., $Icc$) and supply voltage (i.e., $Vcc$), we study two systems: 1) a single-core Coffee Lake (i7-9700K) desktop CPU operating at Turbo frequencies (i.e., $4.9GHz$ and $4.8GHz$), and 2) a two-core Cannon Lake (i3-8121U) mobile CPU operating at Turbo frequencies (i.e., $3.1GHz$ and  $2.2GHz$). We run two workloads on each of the two systems with each of the two Turbo frequencies: 1) a loop that runs scalar instructions, denoted by \emph{Non-AVX}, and 2) a loop that runs \emph{AVX2} instructions.   

\begin{figure}[ht]
\begin{center}
\includegraphics[trim=0.5cm 0.7cm 0.5cm 0.65cm, clip=true,width=1\linewidth,keepaspectratio]{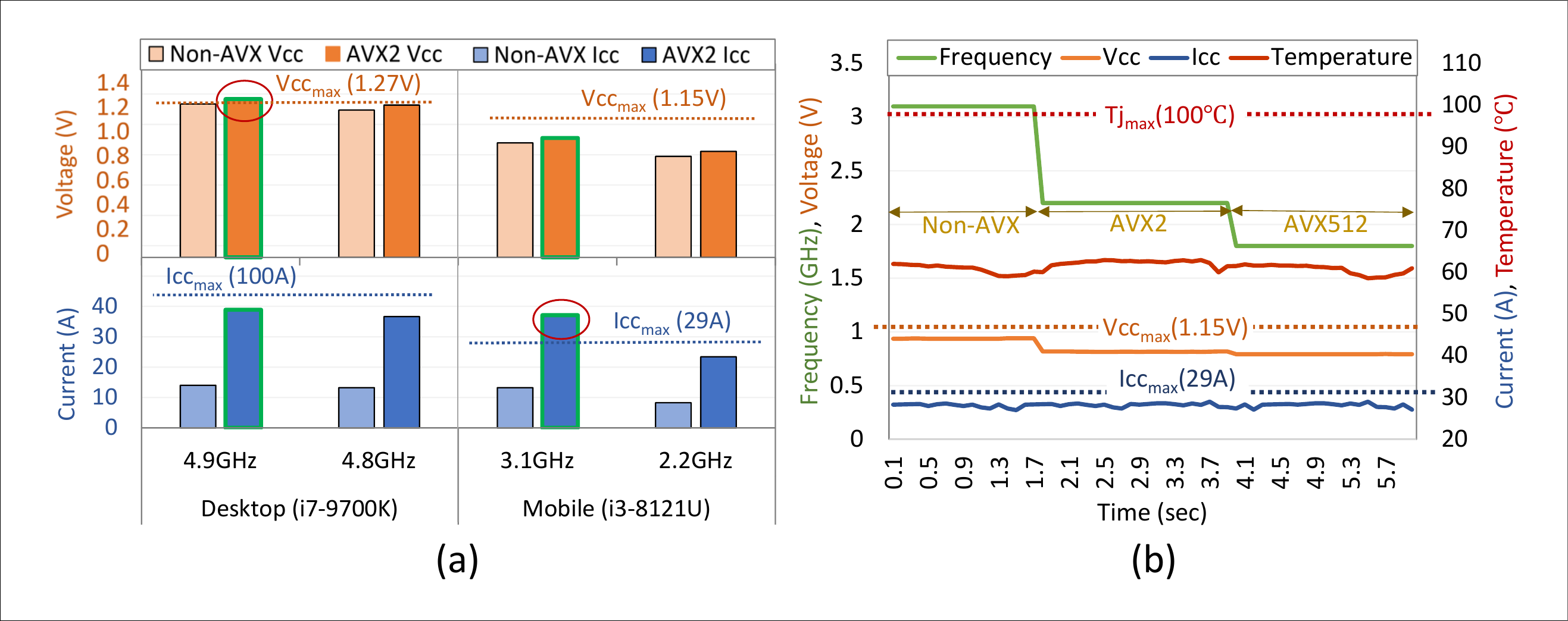}
\caption{(a) $Vcc$ (top), and $Icc$ (bottom) when running non-AVX/AVX2 workloads at two Turbo frequencies on desktop/mobile systems. (b) $Vcc$, $Icc$, core \jk{junction} temperature, and core clock frequency for the mobile system when executing code with three distinct execution phases (Non-AVX, AVX2, and AVX512). $Tj_{max}$ is the maximum allowed \jk{core junction} temperature.}\label{ivccmax}
\end{center}
\end{figure}

Figure~\ref{ivccmax}(a) shows the $Vcc$ (top) and $Icc$ (bottom) measurements from each system when running each workload at each Turbo frequency. All values shown in Figure~\ref{ivccmax}(a) are experimentally measured except the bars with green borders, which are projected \je{values.}\footnote{We extrapolate the current and voltage based on our real system measurements described in Section \ref{sec:method}.}

We make two key observations from Fig~\ref{ivccmax}(a). First, in the desktop system, the supply current consumption when operating at either frequency (\SI{4.9}{\giga\hertz} and \SI{4.8}{\giga\hertz}) is below the system limit ($Icc_{max}=$\SI{100}{\ampere}), while the supply voltage \emph{exceeds} the system limit ($Vcc_{max}=$\SI{1.27}{\volt}) when executing AVX2 code at a frequency of \agythree{\SI{4.9}{\giga\hertz}. In contrast, w}hen the processor operates at \agythree{\SI{4.8}{\giga\hertz}} while running AVX2 code, the supply voltage stays within the $Vcc_{max}$ limit. 
Second, the supply voltage of the mobile system when running AVX2 code at either frequency (\SI{3.1}{\giga\hertz} and \SI{2.2}{\giga\hertz}) is below the system limit ($Vcc_{max}=$\SI{1.15}{\volt}), while the supply current \emph{exceeds} the limit ($Icc_{max}=$\SI{29}{\ampere}) when running AVX2 code at \SI{3.1}{\giga\hertz}. In contrast, when the processor operates at the lower \SI{2.2}{\giga\hertz} frequency while running the same AVX2 code, the supply current stays within the $Icc_{max}$ limit.\footnote{Intel allows exceeding $Icc_{max}$ and $Vcc_{max}$ limits when overclocking a system (e.g., via the BIOS \cite{intel_bios_overclocking} or the XTU tool \cite{intel_xtu_overclocking}). This process is out of the processor's specification and can shorten the processor's lifetime~\cite{intel_bios_overclocking,intel_xtu_overclocking}.}

Fig~\ref{ivccmax}(b) plots the supply current, supply voltage, core \jk{junction} temperature, and core clock frequency for the Cannon Lake mobile system when executing code with three distinct execution phases consisting of instructions with different computational \jk{intensities}: 1) Non-AVX, 2) AVX2, 3) AVX512.

We make three key observations: 1) when the instructions of the current phase require more power (current) than the instructions of a previous phase, the processor reduces its frequency (e.g., $time =$ \SI{1.7}{\second} and \SI{4.0}{\second}) to a level that maintains its supply current below the $Icc_{max}$ limit, 2) the voltage is set to a level corresponding to the new frequency based on the voltage/frequency curves, which is significantly lower than $Vcc_{max}$, and 3) the \jk{junction} temperature (\jk{which is} between 58$^\circ$C and 62$^\circ$C) is \jk{much} lower than the maximum allowed \jk{junction} temperature of the processor, $Tj_{max}$ (100$^\circ$C).   

The Intel architecture provides three Turbo frequency licenses (LVL\{0,1,2\}\_TURBO\_LICENSE) that the processor operates at. \sr{This depends} on the instructions that are being executed and the number of active cores \cite{intel_opt_ref_manual2020}. Kalmbach \emph{et al.}~\cite{kalmbach2020turbocc} exploit this side-effect of throttling the core's frequency when executing PHIs on multiple cores at Turbo frequency to create \jk{a cross-core} covert channel. The authors hypothesize that the side-effect is due to \jk{\emph{thermal}} management mechanisms, while we observe that the side-effect is due to \jk{\emph{current}} management \agy{mechanisms}, which \jk{affect the system even though temperature} is not a problem (as seen in Figure~\ref{ivccmax}(b)). 

We make two major conclusions on the core frequency reduction that follows (within tens of microseconds) the execution of PHIs (e.g., AVX2 and AVX512), based on experimental observations from Figure~\ref{ivccmax}. First, this frequency reduction is mainly due to \jk{the} \emph{maximum instantaneous current limit} (i.e., $Icc_{max}$) and \emph{maximum voltage limit} (i.e., $Vcc_{max}$) protection mechanisms, which keep the processor within its maximum current and maximum voltage design limits when executing PHIs. Second, the frequency reduction is \emph{not} caused by immediate thermal events or thermal management mechanisms, which typically take tens of milliseconds to \jc{tens} of seconds to develop \jk{or react} after an increase in processor power~\cite{rotem2013power,singla2015predictive,rotem2011power,rotem2015intel,brooks2001dynamic}.    

\noindent \textbf{Key Conclusion 2.}
\textit{
Contrary to the state-of-the-art work's hypothesis \cite{kalmbach2020turbocc}, we observe\jk{, on real processors,} that the core frequency reduction that directly follows the execution of PHIs at the maximum (Turbo) frequency is due to \jk{the} maximum instantaneous current limit (i.e., $Icc_{max}$) and maximum voltage limit (i.e., $Vcc_{max}$) protection mechanisms, and not \jk{due to} thermal events or thermal management mechanisms.}


\subsection{AVX Throttling is Not Due to Power Gating}\label{sec:not_power_gating}
\begin{figure}[ht]
\begin{center}
\includegraphics[trim=0.5cm 0.75cm 0.5cm 0.9cm, clip=true,width=1\linewidth,keepaspectratio]{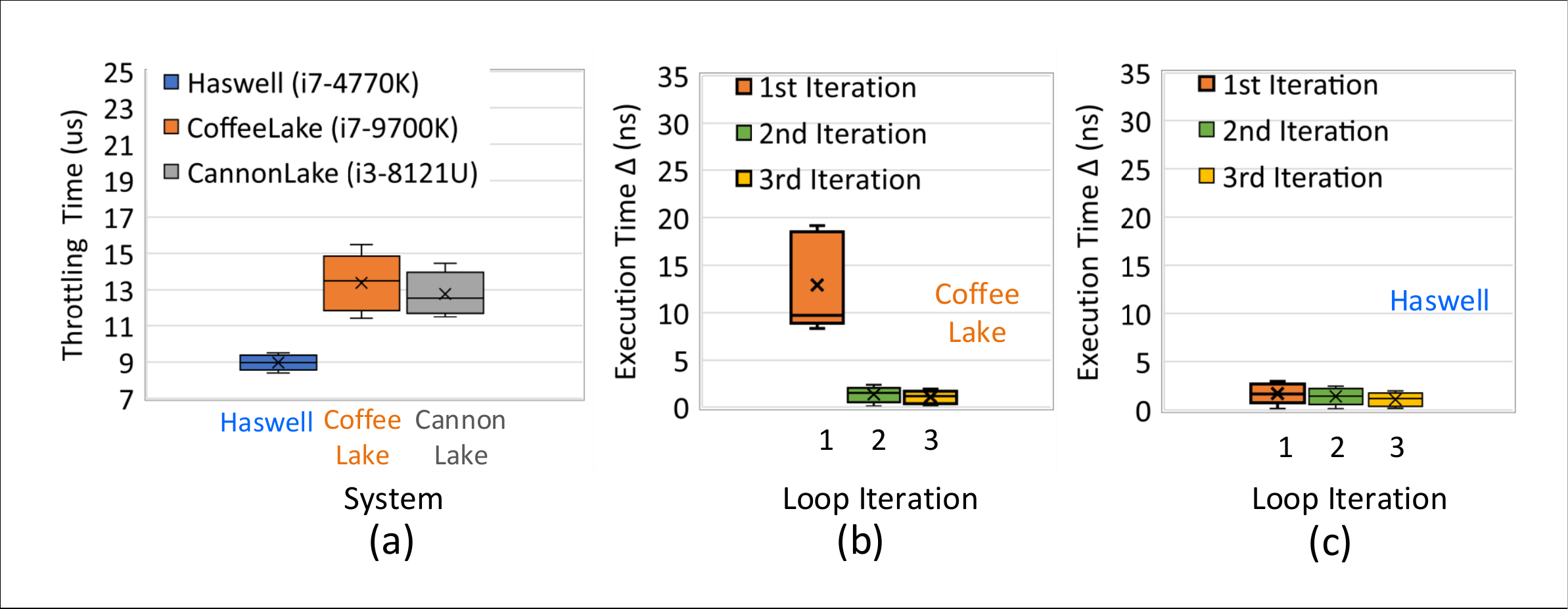}
\caption{(a) Distribution of throttling period for Haswell, Coffee Lake, and Cannon Lake client processors. (b) and (c) show the execution time delta from the expected execution time of AVX2 loop iterations on the Coffee Lake and Haswell processor, respectively.}\label{powergating}
\end{center}
\end{figure}

We next measure the throttling period of AVX2 instructions on three generations of Intel processors (i.e., Haswell, Coffee Lake, and Cannon Lake) to analyze the impact \jk{of the AVX unit's} power-gating on the throttling time of AVX instructions. We plot the distribution of the throttling period for each processor in Fig \ref{powergating}(a). We observe that the throttling side-effect occurs even in the Haswell processor\footnote{The Haswell processor uses a faster voltage regulator (FIVR \cite{2_burton2014fivr}) than Coffee Lake \ja{and} Cannon Lake (MBVR~\cite{haj2019comprehensive}), and therefore has a shorter throttling period.}
(Intel's fourth generation client processor), \emph{although} Intel \jk{reports} that AVX power-gating is a \emph{new} feature \jk{introduced in the later} Skylake processor (Intel's sixth generation client processor) \cite{mandelblat2015technology} to reduce core leakage \jk{power when the AVX unit is \emph{not}} in use. Furthermore, as discussed in Section~\ref{sec:background}, prior works~\cite{kahng2013many,akl2009effective,kahng2012tap} have shown that opening a power-gate takes up to tens of nanoseconds, whereas the throttling period lasts several microseconds. Therefore, we \emph{hypothesize} that the source of throttling is \emph{not} the newly-added power-gating mechanism in the Skylake processor.


To support our hypothesis, we want to find the time it takes to open the AVX2 power-gate by comparing the execution time of multiple subsequent iterations of AVX2 instructions. We track the core clock cycles (CPU\_CLK\_UNHALTED) on the Coffee Lake and Haswell processors when running AVX2 instructions in a loop (consisting of 300 $VMULPD$ instructions that use registers) with a $3GHz$ core clock frequency. \jk{Figures \ref{powergating}(b) and (c) plot the deltas of the execution time (measured with the CPU\_CLK\_UNHALTED performance counter) of the first iteration of the loop (in which the power-gate is being opened) and subsequent iterations (in which the power-gate is already opened)} against the expected execution times (based on operating frequency and instructions' IPC) on Coffee Lake and Haswell processors, respectively, for the first three iterations (all remaining iterations are similar to the third iteration). All iterations run under the throttling side-effect, i.e., at a \emph{quarter} of the expected IPC. 

We observe that the first iteration of the loop running on Coffee Lake (shown in Figure  \ref{powergating}(b)) is more than $8ns$ longer than the other two iterations, while for the Haswell processor (shown in Figure \ref{powergating}(c)), all iterations have nearly the same latency. We conclude that the AVX power-gating feature, implemented in all processors since Skylake, has approximately $8$--\ja{$15$} nanoseconds of wake-up latency, which is only ${\sim}0.1\%$ of the AVX throttling period ($12$--$15$ microseconds, as demonstrated in Fig~~\ref{powergating}(a)). 



\begin{figure}[ht]
\begin{center}
\includegraphics[trim=0.5cm 0.78cm 0.5cm 0.95cm, clip=true,width=1\linewidth,keepaspectratio]{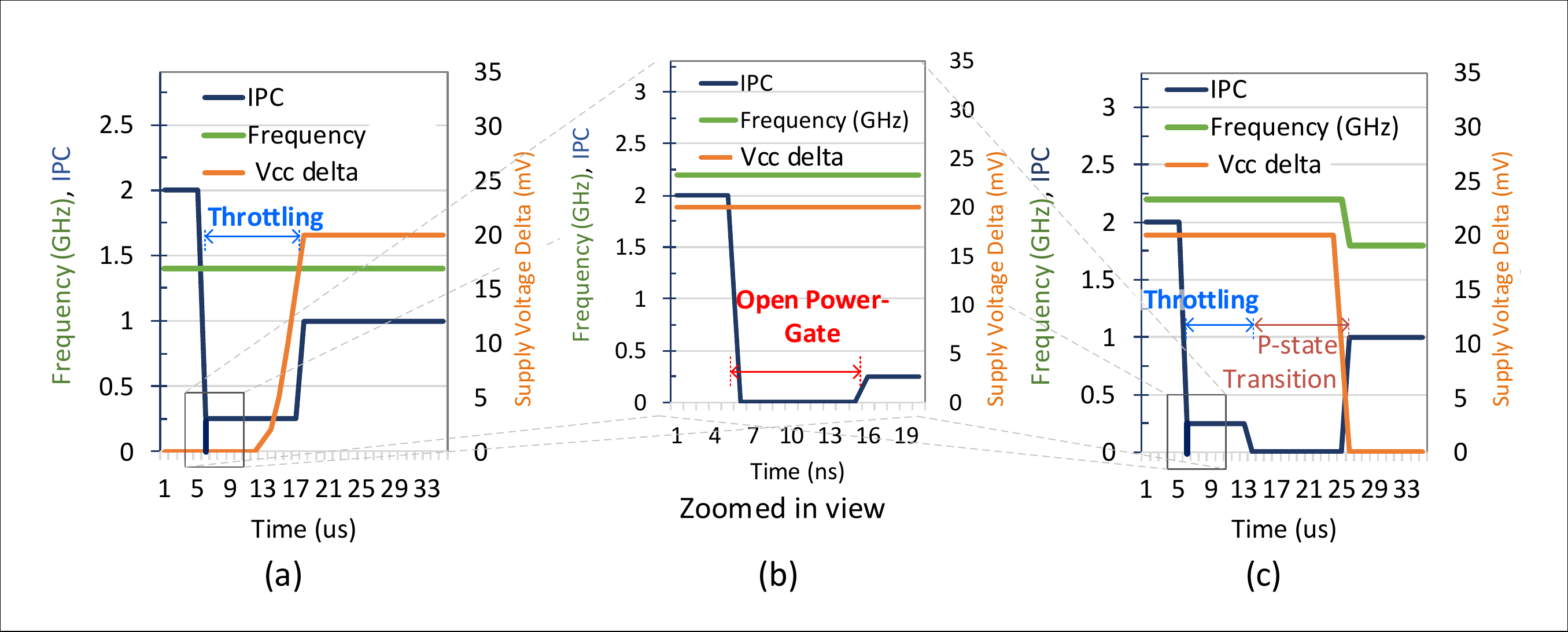}
\caption{\jd{Illustration of the} \ja{AVX power-gate,} $Vcc$, CPU core frequency, and IPC \ja{behavior} \jc{when executing \jkx{the} AVX2 instruction loop that activates two current management mechanisms: (a) $di/dt$ noise prevention by throttling the core execution while increasing the voltage guardband, and (c) $Vcc_{max}$/$Icc_{max}$ design \jd{limit} protection by throttling the core execution while initiating a P-state transition to reduce the voltage and frequency.}  (b) Zoomed in (nanosecond\je{-granularity}) view of the opening of the \ja{AVX2} power-gate. \ja{Core's local PMU opens the AVX2 power-gate when the first AVX2 instruction is executed}.
}\label{pstatetransition}
\end{center}
\end{figure}

\jk{Figure~\ref{pstatetransition} depicts} how AVX power-gating, IPC, frequency, and $Vcc$ changes over time during \jk{the AVX2 PHI execution} \ja{on a Cannon Lake system}. \ja{The} current management mechanisms throttle the CPU core to either 1) increase the $Vcc$ voltage guardband to prevent di/dt noise, (as illustrated \ja{orange plot of} Figure \ref{pstatetransition}(a)), or 2) decrease the core frequency to keep the Vcc/Icc within limits (as illustrated \ja{in orange and green plots of} Figure \ref{pstatetransition}(c)). In the case the power-gate is closed, the processor first opens the power-gate \jk{quickly (i.e.,} within several nanoseconds) and starts executing the PHIs in throttled mode (as shown in Figure \ref{pstatetransition}(b)\jk{, which zooms into the appropriate \jkx{parts} of Figures~\ref{pstatetransition}(a) \ja{and \ref{pstatetransition}(c)})}.

\noindent \textbf{Key Conclusion 3.}
\textit{Contrary to the state-of-the-art work's hypothesis~\cite{schwarz2019netspectre},
power-gating execution units
(e.g., AVX2 power gating \cite{mandelblat2015technology}) 
accounts for an insignificant portion (${\sim}0.1\%$) of the total throttling time observed when executing PHIs.}

\subsection{Multi-level Throttling} \label{sec:crosscore_multi_level}

\begin{figure}[ht]
\begin{center}
\includegraphics[trim=0.5cm 0.75cm 0.5cm 0.9cm, clip=true,width=0.9\linewidth,keepaspectratio]{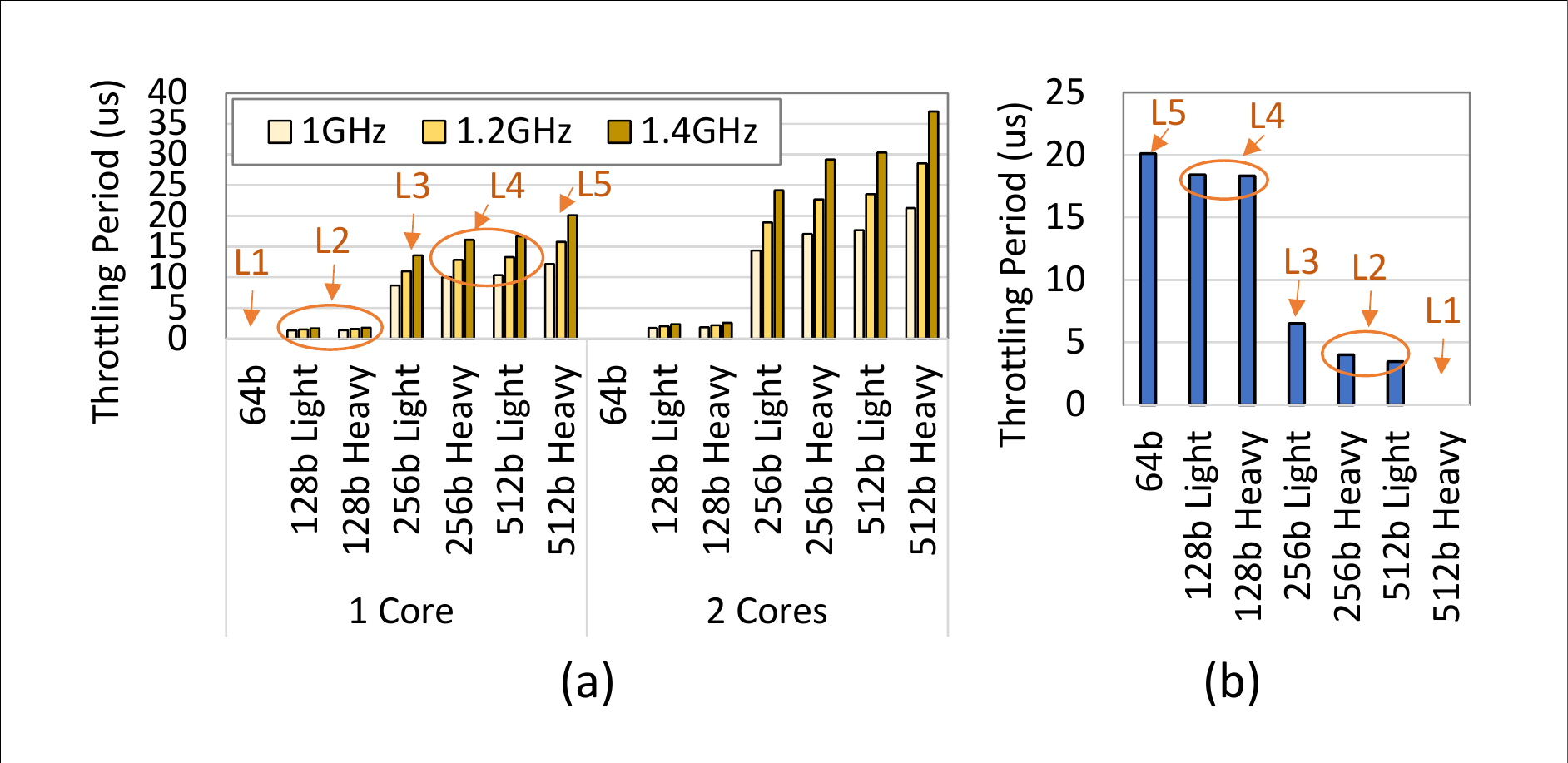}
\caption{ (a) Effect of executing different
instruction types at various frequencies on one and two cores on the length of the throttling period \jk{in the Cannon Lake system}. (b) Throttling period of a 512b\_Heavy loop when \jk{the loop is preceded by} different instruction types (x-axis). Different voltage guardband levels result in a multi-level (L1-L5) throttling period.}\label{multilevelthrottling}
\end{center}
\end{figure}

To further understand the side-effects of \jk{the} throttling technique, we conduct experiments on the \jk{Cannon Lake} processor. We  execute one of seven instruction types (i.e., 64b, 128b\_Light, 128b\_Heavy, 256b\_Light, 256b\_Heavy, 512b\_Light, and 512b\_Heavy, discussed in Section \ref{sec:covert_overview}) in a loop using three experiments. 
In the first and second experiments, we execute the seven instruction types on one and two cores, respectively, at one of three different frequencies ($1GHz$, $1.2GHz$ and $1.4GHz$). We measure the throttling period of each core. Figure~\ref{multilevelthrottling}(a) plots the results of these two experiments.
In the third experiment, we execute the seven instruction types in a loop followed by a loop of 512b\_Heavy (the heaviest instruction type). We measure the throttling period of the 512b\_Heavy loop. 
Figure~\ref{multilevelthrottling}(b) plots the results of the third experiment with \jk{an} operating frequency of $1.4GHz$.

We make two observations from Figure \ref{multilevelthrottling}\crj{(a)}. First, the throttling period increases as the 1) computational intensity (i.e., heaviness \jd{of the operation type} \jc{and/or width}) of the executed instructions increases, \jc{2)} frequency of the core increases, and \jc{3)} number of cores that are concurrently-executing PHIs increases. Increasing any of these parameters increases current consumption, thereby increasing the supply voltage guardband (i.e., $\Delta V$), as we explain in Section \ref{sec:background} (Equation \ref{eqn:guardband}).
Second, when running PHIs on two cores, the throttling period increases significantly (e.g., 256b\_Heavy throttling period is $9us$ on two cores at $1GHz$ compared to $5us$ in \jk{only one} core). The longer throttling period is due to the higher voltage guardband, which requires more time for the processor to increase the voltage to the required level. The processor PMU stops throttling the cores once the shared VR is settled at the required level by \emph{both} cores.  

We observe from Figure \ref{multilevelthrottling}\crj{(b)} that the throttling period of the 512b\_Heavy loop increases when the computational intensity of the instructions \jk{executed in the preceding loop (shown on the x-axis)} decreases. This is because the lower the computational intensity of the instructions in \jk{the preceding} loop, the lower the voltage guardban\jk{d, and thus,} executing the 512b\_Heavy loop requires more time for the processor to increase the voltage to the required level.

We observe from Figures \ref{multilevelthrottling}(a) and (b) that there are at least \emph{five} \jk{throttling} levels (L1–L5) \jk{corresponding to the computational intensity of} instruction types, as shown Figure \ref{multilevelthrottling}(a) and \ref{multilevelthrottling}(b) \jk{(indicated by the orange text)}.

We conclude that 1)~the throttling period of a CPU core when running PHIs has \jk{at least} five levels,\footnote{These five \agythree{throttling} levels, which affect the core throttling period after executing PHIs at any frequency (even if Turbo frequencies are disabled), are different from the three Turbo licenses (i.e., LVL\{0,1,2\}\_TURBO\_LICENSE) discussed in Section \ref{sec:iccmax_vccmax}.} which depends on multiple parameters such as frequency, voltage, and the computational intensity of the instructions, and 2)~there is  \cj{a} \emph{cross-core} throttling side-effect due to \jk{the shared voltage regulator between cores}.

\noindent \textbf{Key Conclusion 4.}
\textit{Current management \agy{mechanisms} result in a multi-level throttling period depending on the computational intensity of the PHIs. This multi-level throttling period has \cj{a}  cross-core side-effect, in which the \agythree{throttling period depends on how many cores are running PHIs.}}

\subsection{Throttling Affects SMT Threads} \label{sec:crossthread_multi_level}
\begin{figure}[ht]
\begin{center}
\includegraphics[trim=0.5cm 0.75cm 0.5cm 0.95cm, clip=true,width=1\linewidth,keepaspectratio]{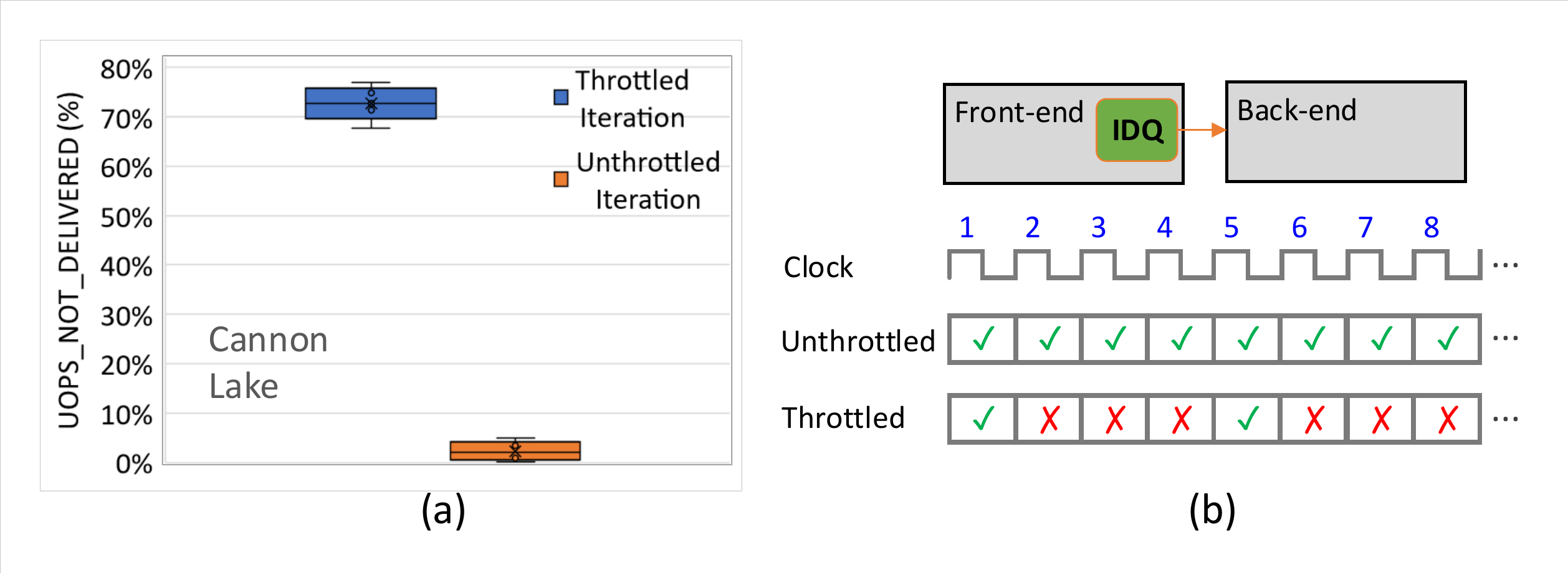}
\caption{(a)~\ja{Normalized} IDQ\_UOPS\_NOT\_DELIVERED performance counter during \ja{throttled and unthrottled} iterations. (b)~Pipeline behavi\agythree{o}r during throttling.}\label{uops_undelivered}
\end{center}
\end{figure}

To understand the source of throttling and its microarchitectural impact on the core, we track the number of micro-operations (uops) that the core pipeline delivers from the front-end to the back-end during throttled and non-throttled loop iterations when running AVX2 PHIs on the Cannon Lake processor. 
To do so, we read the \texttt{IDQ\_UOPS\_NOT\_DELIVERED} and \texttt{CPU\_CLK\_UNHALTED} performance counters at the beginning and end of each iteration. \texttt{IDQ\_UOPS\_NOT\_DELIVERED} counts the number of uops \emph{not delivered} by the Instruction Decode Queue (IDQ) to the back-end of the pipeline when there were \emph{no back-end stalls} \cite{intel_pmons}. We \jb{\emph{normalize}} the counter by the maximum possible number of uops that can be delivered during the iteration, i.e., \texttt{UOPS\_NOT\_DELIVERED} $=$ \texttt{IDQ\_UOPS\_NOT\_DELIVERED/$($4$\cdot$CPU\_CLK\_UNHALTED$)$}.
Figure \ref{uops_undelivered}(a) shows the distribution of the normalized undelivered uops for throttled and non-throttled loop iterations. We observe from Figure \ref{uops_undelivered}(a) that for a throttled loop iteration, the IDQ does \emph{not} deliver any uop in approximately three-quarters (${\sim}75\%$) of the core cycles even though the back-end is \agythree{\emph{not}} stalled. On the other hand, for a non-throttled loop iteration, we observe that there \agythree{are} almost \ja{\emph{no}} cycles in which the IDQ does not deliver uops to the back-end.

We conclude that the core uses a throttling mechanism that limits the number of uops delivered from the IDQ to the back-end during a certain time window, as illustrated in Figure~\ref{uops_undelivered}(b). During a time window of four core clock cycles, the IDQ delivers uops to the back-end \ja{in} only one cycle, while \ja{in} the remaining three cycles, the throttling mechanism blocks the IDQ (i.e., no uops are delivered). We found that this throttling mechanism affects \emph{both threads in Simultaneous \ja{Multi-Threading}} (SMT). 
This means that both threads are throttled when one thread executes \agythree{a} PHI because the IDQ\agythree{-to-}back-end interface is shared between the threads. 


\noindent \textbf{Key Conclusion 5.}
\textit{Contrary to the \agythree{hypothesis of a} state-of-the-art \agythree{work}~\cite{schwarz2019netspectre}, we observe on a real system that 
\agythree{the processor front-end (IDQ)}
to back-end \agythree{uop delivery} is blocked during 75\% of the time, rather than a reduced core clock frequency of 4$\times$. This throttling mechanism affects both threads in Simultaneous \ja{Multi-Threading} (SMT).}





\subsection{Software-level Power Management Policies}

To examine whether software-level power management policies affect the underlying mechanisms of IChannels, we observe the underlying mechanism under three separate power management \jkx{policy} \ja{governors \cite{pallipadi2006ondemand}}: userspace, powersave, \ja{and} performance. We find that the underlying mechanism of IChannels persists across all three policies. We also find that software-level power management policies do not affect the hardware throttling mechanisms. This is because hardware throttling is implemented inside the core for fast response (i.e., nanoseconds), and we are not aware of any software control that allows to disable this mechanism.

%% file: _06_results.tex
\section{\tech Evaluation }
\label{sec:eval}

This section evaluates our three {\tech} covert channels, \ta,  \tb, and \tc, using our proof-of-concept (PoC) implementations.  
We compare our three {\tech} covert channels to recent works~\cite{schwarz2019netspectre,alagappan2017dfs,kalmbach2020turbocc,khatamifard2019powert} that exploit PHIs and power management vulnerabilities \ja{\jb{to create} covert channels}. 
\juang{In this evaluation, we} build covert channels between execution contexts running on: \juang{(1)} the same hardware thread, \juang{(2)} the same physical core but \ja{using} different SMT threads, \juang{and (3)} different physical cores.

\subsection{Setup}
We evaluate \tech{} on 
\juang{Coffee Lake (Core i7-9700K \cite{coffeelake_2020}) and Cannon Lake (Core i3-8121U \cite{i38121u_cannonlake}).}
We test \ta and \tc on both 
processors, but \sr{we test} \tb only on  
\juang{Cannon Lake} \ja{as} 
\juang{\ja{Coffee} Lake} 
does not support SMT.

\subsection{{\tech} Throughput}\label{sec:throughput}
The throughput \juang{(i.e., the channel capacity)} of our three \jb{\tech} side channels is \jkx{${\sim}2.9Kbps$}. 
Each channel send\agy{s} $2$ bits \jb{in} each \ja{communication transaction} cycle. The actual sending of the $2$ bits takes less than $<40us$, during which the \texttt{Sender} code executes the appropriate PHI while the \texttt{Receiver} code measures the throttling-period (TP) to decode two transmitted bits. Before sending the next two bits, each covert channel should wait for \emph{reset-time} (${\sim}650us$), as we discuss in Section~\ref{sec:IccThreadCovert}. 
Therefore, each covert channel's cycle time is the sum of \juang{the} latencies for \ja{reset-time} and transmitting two bits, which is less than $690us$. Next, we compare each covert channel's throughput to its \juang{most} relevant prior work.

\begin{figure}[h]
\begin{center}
\includegraphics[trim=0.6cm 0.7cm 0.6cm 0.7cm, clip=true,width=0.85\linewidth,keepaspectratio]{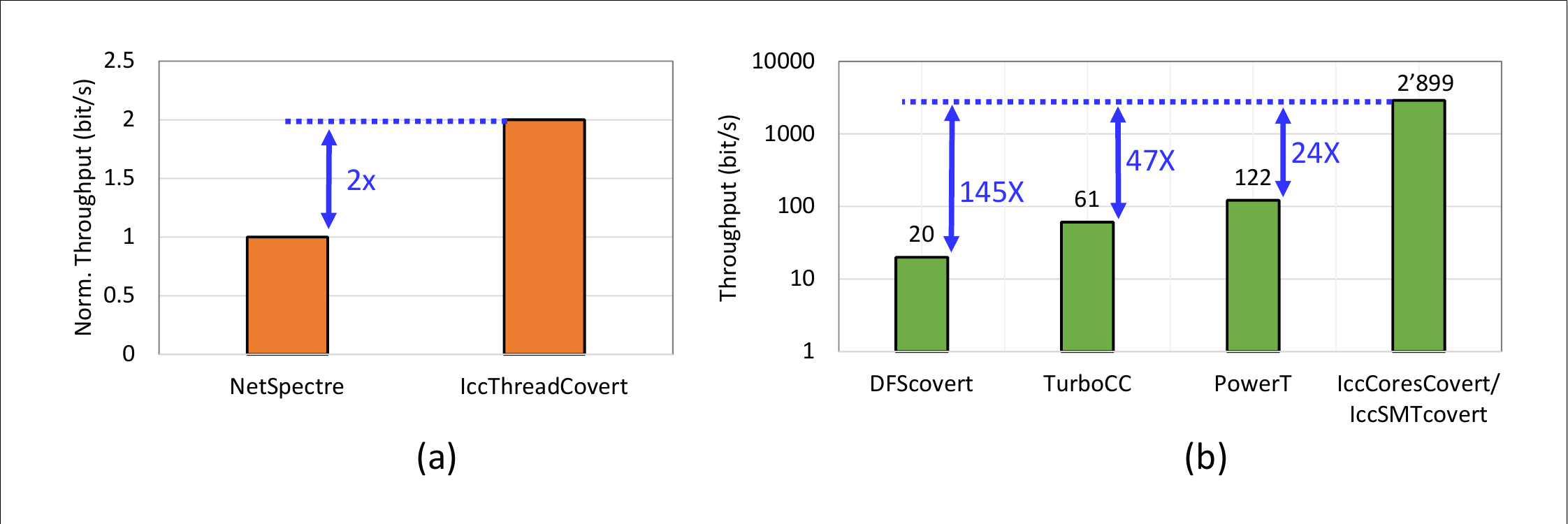}
\caption{(a) \ja{Normalized} throughput of \ta  and NetSpectre~\cite{schwarz2019netspectre}. (b) \ja{Throughput of} \tb, \tc, DFScovert \cite{alagappan2017dfs}, TurboCC \cite{kalmbach2020turbocc} and PowerT \cite{khatamifard2019powert}.} \label{fig:evaluation}
\end{center}
\end{figure}

\noindent \textbf{\ta.}
\agy{We compare \ta against NetSpectre~\cite{schwarz2019netspectre}, the state-of-the-art \jc{work}
\jc{that exploits the variable latency of PHIs to create a covert channel between} two execution contexts running on the \emph{same hardware thread}.}
\juang{Similarly} to \ta, NetSpectre exploits the throttling of AVX2 instructions to establish a covert channel within the same hardware thread. The NetSpectre covert channel can send \emph{one bit} per communication \ja{transaction}, while the \ta covert channel can send \emph{two bits} \ja{per communication transaction}. Figure \ref{fig:evaluation}(a) \juang{compares} \jb{the} \ta throughput to \juang{that of} NetSpectre. 
\juang{Our results show that \jb{the} \ta throughput is $2\times$ that of NetSpectre.\footnote{\crj{
\ja{We compare \ta to NetSpectre\jb{'s} main gadget (i.e., throttling of PHIs on the same hardware thread), and not to the end-to-end NetSpectre implementation\jc{, which includes network components}.}}}}


\noindent \textbf{\tb and \tc.} 
\ja{The three most related works} to \tb and \tc, which create covert channels across \juang{SMT threads and across cores}, are DFScovert \cite{alagappan2017dfs}, TurboCC \cite{kalmbach2020turbocc}, and PowerT \cite{khatamifard2019powert}. 
These works exploit different power management mechanisms of modern processors to build covert \juang{channels} across cores and SMT threads. DFScovert manipulates the power governors that control the CPU core frequency. TurboCC utilizes CPU core frequency changes due to the 
\juang{frequency boosting} mechanism of modern Intel processors (i.e., Intel Turbo \cite{rotem2012power,rotem2013power}). PowerT utilizes CPU core frequency changes due to the thermal management \juang{mechanisms} in modern processors. These three mechanisms are slow (\juang{i.e.,} can take several milliseconds) compared to the current management mechanisms that our \tech covert channels utilize. Figure \ref{fig:evaluation}(b) compares the throughput of our \tb and \tc  mechanisms to DFScovert, TurboCC, and PowerT. \tb/\tc throughput is  $145\times$ \jb{($2899/20$)}, $47\times$ \jb{($2899/61$)}\ja{,} and $24\times$ \jb{($2899/122$)} the throughput of DFScovert, TurboCC\ja{,} and PowerT, respectively.      


\subsection{\ja{System Noise Effect on} {\tech} \ja{Accuracy}}

\jh{Similar to all recent covert channels (e.g., Spectre \cite{kocher2019spectre}, Meltdown \cite{lipp2018meltdown}, NetSpectre \cite{schwarz2019netspectre}, DFScovert \cite{alagappan2017dfs}, TurboCC \cite{kalmbach2020turbocc}, and PowerT \cite{khatamifard2019powert}), {\tech} covert channels\jb{'} \ja{accuracy is} sensitive to \ja{system} noise. \ja{System} \juang{noise}  can occur in two main scenarios. First, due to system activity, such as interrupts and context switches, which can extend the execution time measured by the 
\juan{\texttt{Receiver}}, 
\juan{causing} errors in decoding \jb{of} the received bits. 
Second, an application running concurrently with the covert channel's \texttt{Sender}/\texttt{Receiver} processes can reduce the covert channel's accuracy by, for example, executing PHIs.
In this section, we analyze the {\tech} error rate \jb{under} these \jb{two} scenarios and \ja{propose} approaches to mitigate the effects of noise.}

\begin{figure}[h]
\begin{center}
\includegraphics[trim=0.6cm 1.1cm  0.6cm 0.95cm, clip=true,width=0.95\linewidth,keepaspectratio]{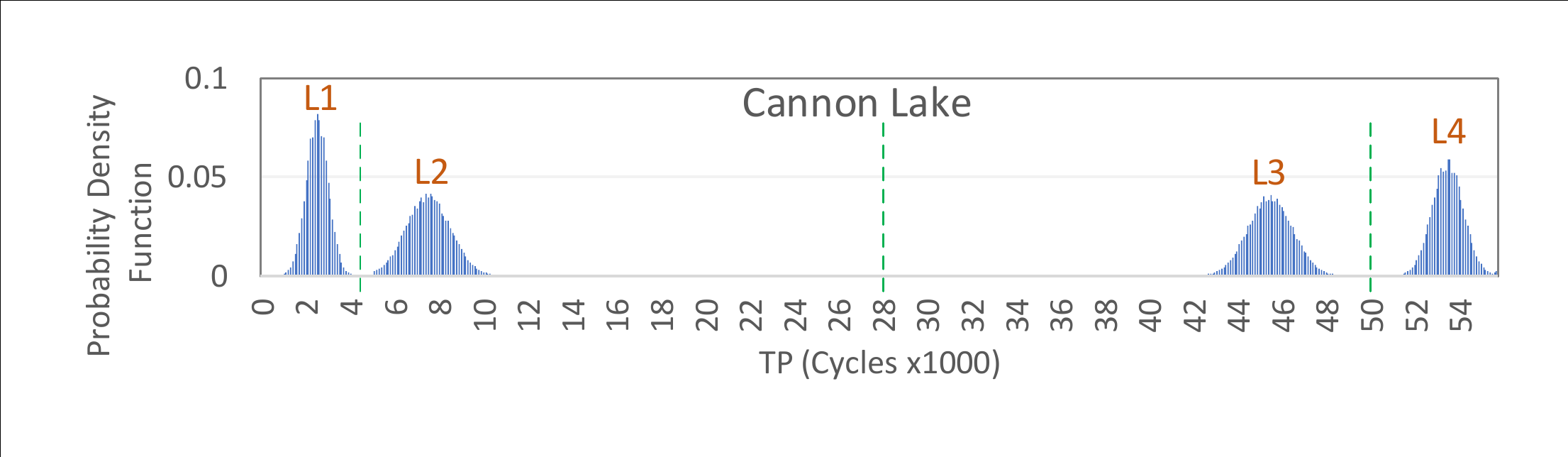}
\caption{Distribution of \ja{throttling period} (TP) values in the \texttt{Receiver} when detecting each of the four \juang{throttling} levels \ja{(L1-L4)} \jb{in a low-noise system}. 
} \label{distribution}
\end{center}
\end{figure}

\noindent \textbf{\jb{Low-Noise} System.} 
\jh{Figure \ref{distribution} shows the distribution of the throttling period (TP) counter (in cycles) of the \texttt{Receiver} code for a client system with \jb{relatively low} \crj{noise} 
(\juan{i.e.,} interrupt and context-switch rates below $1000$ events per second) while other \emph{non-AVX} applications (besides the \juan{\texttt{Sender/Receiver}}) are running on the system.
The TP values are centered around the threshold of each of the four-level ranges \crj{marked in the green  dashed line in Figure \ref{distribution}}. The ranges do not overlap since there is a significant difference in the values ($>2K$ cycles). Therefore, the error rate of the \jkx{\tech covert} channel is close to zero when there is \jb{low} noise in the system.}

\begin{figure}[h]
\begin{center}
\includegraphics[trim=1.25cm 0.6cm  0.9cm 0.6cm, clip=true,width=0.95\linewidth,keepaspectratio]{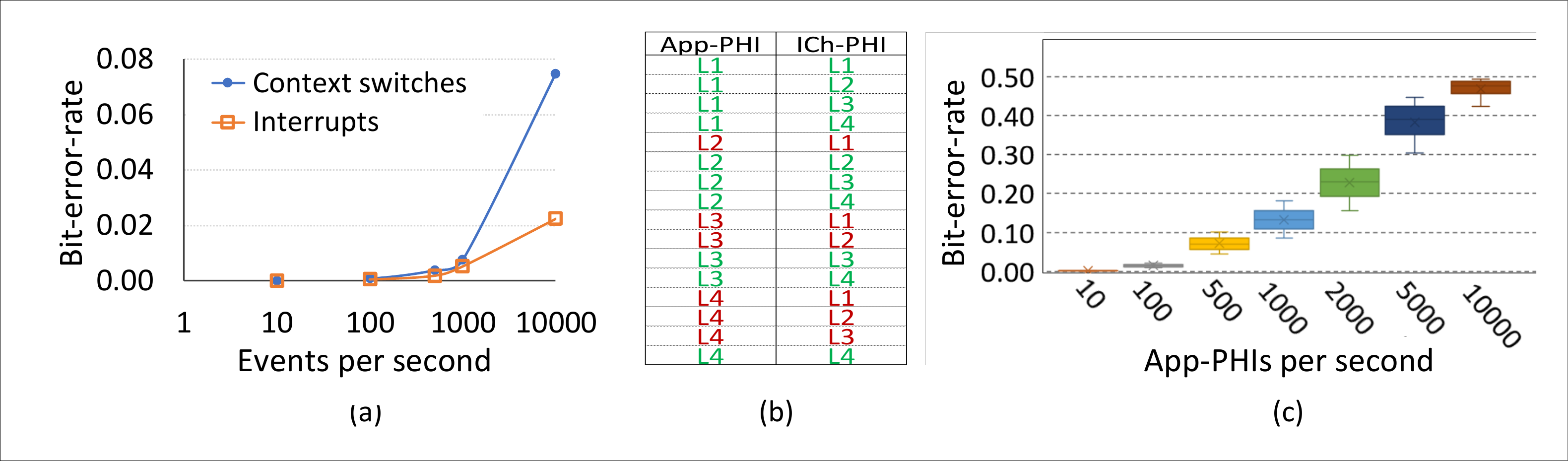}
\caption{(a) \ja{Bit-error-rate} (BER) as a function of system event (context-switch or interrupt) rates. (b) \crj{Cases of erroneous decoding (red)} \ja{due to a noise that occurs} when an application executes PHIs (App-PHI) concurrently \jkx{with the} \crj{\tech} \texttt{Sender}/\texttt{Receiver} processes \jkx{that execute} PHIs \crj{(ICh-PHI)}. (c) Distribution of BER when running  a synthetic Application  (App) concurrently with the \texttt{Sender}/\texttt{Receiver}, which \jb{executes} PHIs \jb{at} different rates. 
} \label{cs_int_ber}
\end{center}
\end{figure}

\noindent \textbf{\jb{High-}Noise from Interrupts and Context \juang{Switches}.} 
\jh{
Interrupts and context-switches that occur while the \texttt{Receiver} is decoding  the received bits (i.e., measuring the execution time of instructions in green in Figure \ref{coverts}) can significantly increase the measured time (on 
\juan{the order} of several microseconds), thereby 
\juan{causing} errors in decoding.
Interrupt and context-switch latencies are typically within few microseconds \cite{he2018sgxlinger,van2018nemesis} to few tens of microseconds \cite{gravani2019iskios,weaver2013linux}, respectively, with occurrence frequencies of a few hundred (e.g., in a 
\juang{noisy} system) to thousands (e.g., in a 
\jb{highly noisy} system) \jb{of events per second}. Figure \ref{cs_int_ber}(a) shows the measured \ja{bit-error-rate} (BER) as a function of system event 
\juan{(i.e., context-switch or interrupt) occurrence} rates. The results show that the BER is low even when the system is highly 
\juang{noisy}. This is because these events have a low probability of occurring during the 
\juang{decoding} interval (\jkx{only several} microseconds) \juang{at the \texttt{Receiver}.} 
}

\noindent \textbf{Noise from \juang{Concurrent} Applications.} 
\jh{An application that 
\juan{executes} PHIs while running concurrently with the \jb{\tech'} \texttt{Sender}/\texttt{Receiver} processes can cause bit errors \jkx{in the covert channel}. In particular, the error can occur if the power level of the application's PHI is higher than that of {\tech} PHIs executed by the \texttt{Sender}/\texttt{Receiver} processes as shown in Figure \ref{cs_int_ber}(b) (in red). Figure \ref{cs_int_ber}(c) shows the measured distribution of \jkx{channel} BER when running a synthetic application (App) concurrently with the IChannels \juan{processes}. 
App injects 
PHIs with a random power level (\jb{from} the four levels) using different  rates ($10$--$10,000$ \ja{App-PHIs per second)}. The figure shows that the BER significantly increases when App executes PHIs at a higher rate.

We evaluate the effect of concurrently running the 7-zip \cite{app_7_Zip} application (which uses AVX2 instructions but not AVX-512) and observe a BER of less than $0.07$ across the three {\tech} covert channels when sending data \ja{during} $60$ seconds.}

\noindent \textbf{Mitigating the Effects of \jb{System}  Noise.}
\jh{We discuss three approaches to mitigating the effects of \jb{system} noise on covert channels. First, an attacker can send the secret value many times and \juang{the \texttt{Receiver} can} average the received value over a large number of measurements to obtain the secret value with \je{an} acceptable 
\juang{error rate}.
\juang{Second, we can} use error detection and correction codes. This approach is used by several recent covert channel works~\cite{maurice2017hello,giechaskiel2020c,sehatbakhsh2020new,Guri2020,kalmbach2020turbocc}. 
\juang{Third, the \texttt{Sender/Receiver} can}  initiate \ja{a communication transaction} \juang{only} during \jb{low-noise} periods \cite{ritzdorf2012analyzing}. In case of {\tech}, the sender can track the system status and only transmit data when no other PHIs are executing. Client processors (i.e., \jkx{our main} target \je{systems}) are often idle ($>80\%$ of the day~\cite{stedman2005reducing,haj_connected_standby,comscore2017}) due to low utilization, which provides many opportunities for the sender to reliably transmit data with \juang{little or no} \jb{system} noise.}

\subsection{\tech on Server Processors}\label{sec:otherprocessor}

We have already shown that \emph{all} Intel client \hg{and server}\footnote{\hg{An Intel CPU core has nearly the same {microarchitecture} for client and server \agyfour{processors}. Intel \agyfour{CPU} core \agyfour{design} is a single development project, \agyfour{leading to} a master superset core. The project} \jkt{has} \hg{two derivatives, one for server and one for client processors~\cite{Skylake_die_server}.}} processors from the last decade (\juan{from} \ja{Sandy Bridge} client, 2010, to \ja{Ice Lake} client, 2020  \hg{and \ja{Sandy Bridge} server, 2011, to  \ja{Cascade Lake} server, 2019)}, which can be found in hundreds of millions of 
\juan{devices} in use today \cite{intel_shipped_processors}, are affected by \jb{at least one of} our three proposed covert-channels. 

\subsection{\ja{Side-channel} Attacks Exploiting \tech Side-effects}
\ho{\mbox{\tech} throttling side-effects (e.g., Multi-Throttling-SMT and Multi-Throttling-Cores) can \jb{also} be used as a side-channel to leak data} \jkt{from victim code}. \ho{Attacker code can infer the} \jkt{instruction types} \ho{(e.g., 64bit scalar, 128bit vector, 256bit vector, 512bit vector instructions)} \jkt{of victim} \ho{code that is running 1) on 
\juang{another} SMT thread by utilizing the Multi-Throttling-SMT side-effect, or 2) on another core by utilizing the Multi-Throttling-Cores side-effect.  
In fact, our \jb{proof-of-concept} code of the three \mbox{\tech} covert channels can be used, with minimal changes, to demonstrate a synthetic side-channel between} \jkt{attacker code and victim code}. \ho{However, it would be challenging and requires significant additional effort to conduct a real-world attack\ja{,} which can extract sensitive information using the leaked data (i.e.,} \jkt{the instruction  type} \ho{executed by the victim code). \juang{We}} \jkt{leave} \ho{such side-channel attacks} \jkt{to} \ho{future work.}

%% file: _07_mitigation.tex
\section{Mitigations for \tech} \label{sec:mitigations}

In this section, we propose practical \agyfour{hardware or software techniques} for the mitigation of \tech covert channels.

\noindent \textbf{Fast Per-core Voltage Regulators.} 
As explained in Section \ref{sec:MultiThrottlingCores}, the {largest} side-effect that {\tech} exploits is the \agyfour{c}ross-core side-effect\agyfour{, which} occurs when two cores execute PHIs simultaneously {(\tc)}. When this happens, the throttling period of one core is extended because the other core also requires a voltage increase (to execute PHIs at the same time). To eliminate this \agyfour{c}ross-core side-effect, we propose to implement a PDN with \emph{per-core} voltage regulators\jk{,} such as \lowercase{\emph{Low Dropout Voltage Regulators}} (LDO  \cite{singh20173,singh2018zen,burd2019zeppelin,beck2018zeppelin,toprak20145,sinkar2013low}) or \lowercase{\emph{Integrated Voltage Regulators}} (IVR  \cite{2_burton2014fivr,5_nalamalpu2015broadwell,tam2018skylake,icelake2020}). 
Doing so allows each core {that executes} PHIs to handle its individual voltage transitions using its dedicated VR. 
    

On the other hand, to mitigate \ta and \tb, a fast voltage ramp is required\agyfour{. This is because} the root cause of the relatively long throttling period ($>10us$) is the long time it takes for the VR to increase the voltage, while the throttling due to power-gating is {only} approximately $15ns$ (${\sim}0.1\%$ of the throttling time), as explained in Section \agyfour{\ref{sec:not_power_gating}}. We also show {in Section \ref{sec:not_power_gating}} that processors with IVR, such as Haswell \cite{2_burton2014fivr}, still have \agyfour{a} long \agyfour{throttling period} (e.g., ${\sim}9us$), \agyfour{even though} the IVR PDN has a faster voltage ramp time relative to the motherboard voltage regulators (MBVR \cite{rotem2011power,10_jahagirdar2012power,11_fayneh20164,12_howse2015tick}) used in {some of the systems we evaluate} \agyfour{(i.e., Coffee Lake and Cannon Lake)}. To mitigate this issue, we propose to implement an LDO PDN, which exi\agyfour{s}ts in most recent AMD processors \cite{singh20173,singh2018zen,burd2019zeppelin,beck2018zeppelin,toprak20145,sinkar2013low}). The LDO PDN allows fast voltage transitions (e.g., $<0.5us$) \cite{luria2016dual}, \agyfour{thereby} significantly reduc\agyfour{ing} the throttling time. Reducing the throttling time from $>10us$ to $<0.5us$ does not \emph{completely} \jkx{eliminate} the covert channel. \agyfour{However,} it makes establishing such a covert channel \je{is} much \agyfour{more difficult.}

\jkt{\ja{The} LDO PDN \jkx{incurs} an area overhead of $11\%-13\%$ \cite{kim2013reducing,quelen2018ldo} of the core. However, \ja{this} PDN \ja{can} significantly benefit the system by reducing board area (e.g., by reducing the number of off-chip VRs), enabling fast voltage transitions, and improving energy efficiency \cite{haj2020flexwatts,haj2019comprehensive}.} 

\noindent \textbf{Improved {Core} Throttling.} The \tb covert channel exploits the throttling side-effect that we illustrate in Figure \ref{uops_undelivered}. Once a thread executes a PHI, the processor aggressively throttles the entire core, i.e., blocks delivering uops from the front-end to back-end {for three cycles in a window of four cycles, which affects \emph{both threads}} in an SMT core.  We propose to implement a more efficient throttling mechanism in two stages. First, instead of blocking all the uops sent from the front-end (IDQ) to {the backend, the core blocks} only the uops {that} belong to the thread that executes the PHI. Second, {our enhanced} throttling mechanism {does} not block non-PHI uops.

{This mitigation option should not incur area, power, or performance overhead if} \jkt{carefully} implemented{.} \agyfour{I}n fact, it would likely improve performance by increasing instruction-level concurrenc\agyfour{y}. {However}, it may add more implementation\agyfour{,} design\agyfour{,} and verification effort.

\noindent \textbf{A \agyfour{New} Secur\agyfour{e M}ode of Operation.} While current management \agy{mechanisms} \ja{can improve system energy efficiency}, we have shown that these \agy{mechanisms} can result in security vulnerabilities in the system. To avoid these vulnerabilities, especially when handling sensitive data, 
we propose to add a new mode of system operation to the processor: \emph{secure-mode}. The user can enable secure-mode at any point in time, during which the processor begins to operate at its highest voltage guardband corresponding to the worst-case power virus. While in secure-mode, throttling does not occur as a result of increased power requirements (e.g., executing PHIs) since the voltage guardband is already set to the maximum {level}.  
While secure-mode results in increased power consumption \hy{(by up to $4\%/11\%$ for {a} system with AVX2/AVX512 due to {the} additional voltage guardband)},
it can significantly improve the security of the overall system by mitigating the three covert channels of \tech. 
\agyfour{In contrast, existing security techniques incur significant performance and energy overheads.}
\hy{For example, the well-known Intel SGX \agyfour{mechanism} can incur significant performance degradation (e.g., up to $79\%$ \mbox{\cite{weisse2017regaining}}) and an increase in energy consumption (up to $67\%$ \mbox{\cite{gottel2018security}}) compared to running {in native (i.e., non-SGX)} mode.} \jkt{In addition, SGX increases the area of the memory controller since it requires 1) a memory-encryption-engine (MEE \mbox{\cite{gueron2016memory}}) and 2) additional area to maintain SGX metadata}~\cite{gueron2016memory}.

\hy{Table \mbox{\ref{tab:mitigations}} summarizes the effectiveness and overhead of our proposed mitigation techniques {with respect to} each one of the three covert channels of \mbox{\tech}.}

\vspace{5pt}
\begin{table}[!ht]
\centering
\resizebox{\linewidth}{!}{%
\begin{tabular}{|l||c|c|c|c|}
\hline
\textbf{Mitigation} & \textbf{\ta} & \textbf{\tb} & \textbf{\tc} & \textbf{Overhead}        \\ \hline
\hline

Per-core VR         & Partially     & Partially     & \cmark        & 11\%-13\% more area      \\ \hline
Improved Throttling  & \xmark        & \cmark        & \xmark        & Some design effort       \\ \hline
Secure-Mode         & \cmark        & \cmark        & \cmark        & 4\%-11\% additional power \\ \hline
\end{tabular}%
}
\caption{\hy{Effectiveness and overhead of our mitigation techniques}}
\label{tab:mitigations}
\end{table}

\noindent \textbf{\tech on other Microarchitectures. }
The greater part of a processor's microarchitecture is usually not publicly documented. Thus, it is virtually impossible to know the differences in the implementations (e.g., different throttling techniques, power delivery, resource sharing in SMT, current management \agyfour{at the} circuit level) that allow or prevent {\tech} in other processor architectures (e.g., AMD or ARM) without very careful examination. 
In fact, we confirmed that naively porting \tech to recent AMD and ARM processors does not work.
However, we believe the insights provided by our study (e.g., the multi-level throttling side-effect due to voltage changes, interference between SMT threads while throttling, interference between cores due to sharing {of} the voltage-regulator) on Intel processors can be applied to effectively adapt {\tech} to other processors and inspire experts on {other} microarchitectures to perform the necessary studies for revealing similar covert channels.

%% file: _08_related_works.tex
\section{Related \lois{Work}}

 \input{_07_compare_table}

\lois{To our knowledge, this is the first paper that \lois{demonstrates} how the multi-level throttling side-effects of current management \agy{mechanisms} can be used to build covert channels across execution contexts running on the same thread, {across SMT threads, and across cores} in modern processors. We discuss {and compare to} state-of-the-art covert channels~\cite{kalmbach2020turbocc,schwarz2019netspectre} due to throttling \jh{techniques} of current management \agy{mechanisms} \jh{in Table \ref{tab:prior_phi_works1}} {and Section~\ref{sec:eval}}. We also compare {to} state-of-the-art power~\cite{khatamifard2019powert} and frequency covert channels~\cite{kalmbach2020turbocc,alagappan2017dfs} in {Section~\ref{sec:eval}}.} 
{Other recent} works on covert channels tend to follow one of {the below} directions.

\noindent\textbf{\lois{Power Management Vulnerabilities.}} \lois{There are many works that exploit different power management vulnerabilities~\cite{JayashankaraShridevi2016,Tang2017,zhang2018blacklist,alagappan2017dfs,giechaskiel2020c,giechaskiel2020c,sehatbakhsh2020new,Guri2020}. JayashankaraShridevi \textit{et al.}~\cite{JayashankaraShridevi2016} create two attacks that significantly degrade power efficiency by \lois{inserting} a hardware Trojan in the power management unit. Tang \textit{et al.}~\cite{Tang2017} propose an attack that {infers secret keys by enforcing anomalous frequencies and voltages in the DVFS, and inducing timing errors}. 
Alagappan \textit{et al.}~\cite{alagappan2017dfs} create a covert channel between a Trojan and {a spy process} by modulating the processor operating frequency, with the help of power governors, to stealthily exfiltrate data.
\ja{Giechaskiel \textit{et al.}~\cite{giechaskiel2020c}} create a covert channel between independent \ja{FPGA} boards in a data center by using the power supply unit\agyfour{,} without physical access \agyfour{to boards}. 
Compared to these previous works, \tech is the first work that exploits multi-level throttling side-effects of current management mechanisms {to} develop a new set of covert channels}, providing {more than} $24 \times$ higher channel capacity than {the state-of-the-art power management\agyfour{-based} covert channel} works ({see}  Section~\ref{sec:eval}).

{{Also} compared to these previous works, 
\tech} is 1) \agyfour{\emph{faster}} as it \lois{uses} a mechanism with \lois{a response time of {only} a} few hundreds of microseconds (i.e., throttling due to current management \agy{mechanisms}), while \ja{CPU core} thermal and \ja{governor-controlled \cite{alagappan2017dfs,pallipadi2006ondemand}} frequency changes normally \lois{take}  \ja{several} milliseconds, and  2) \lois{more} \agyfour{\emph{robust}} as the processor gives priority to handling issues due to current emergencies (\agyfour{e.g.,} throttling starts in a few nanoseconds) while thermal and frequency changes are considered less urgent and can be handled {on the order of} milliseconds~\cite{alagappan2017dfs,pallipadi2006ondemand,hajsysscale,gough2015cpu,hanson2007thermal}.   

\noindent\textbf{\lois{Other Covert Channels}.} There are many other types of cover\sr{t} channels that \lois{exploit} other components of the system, such as CPU caches~\cite{percival2005cache,wu2014whispers,maurice2017hello}, DRAM row buffers~\cite{pessl2016drama}, CPU functional units~\cite{wang2006covert}\jd{, or the RowHammer phenomenon in DRAM \cite{kwong2020rambleed,kim2014flipping,kim2020revisiting,mutlu2019rowhammer,frigo2020trrespass}}. {None of these exploit current management mechanisms like \tech does.}

%% file: _07_compare_table.tex
\begin{table*}[ht!]
\vspace{3pt}
\footnotesize
\begin{center}
 \vspace{-15pt}
\resizebox{0.85\linewidth}{!}{%
\begin{tabular}{ |c||c|c|c|c|c|c|c|c|c| }
\hline 
      \textbf{Proposal} \rule{0pt}{3ex}
	& \begin{tabular}{@{}c@{}}\textbf{Same} \\ \textbf{Core} \end{tabular}
	& \begin{tabular}{@{}c@{}}\textbf{\agyfour{C}ross-SMT} \\ \textbf{Threads} \end{tabular} 
	& \begin{tabular}{@{}c@{}}\textbf{\agyfour{C}ross-} \\ \textbf{Core}\end{tabular} 
	& \textbf{BW}
	& \begin{tabular}{@{}c@{}}\textbf{User/}\\\textbf{Kernel}\end{tabular}
	& \begin{tabular}{@{}c@{}}\textbf{Underlying}\\\textbf{Mechanisms}\end{tabular} 
	& \begin{tabular}{@{}c@{}}\textbf{Turbo-} \\ \textbf{Independent}\end{tabular} 
	& \begin{tabular}{@{}c@{}}\textbf{Root} \\ \textbf{Cause}\end{tabular} 
	& \begin{tabular}{@{}c@{}}\textbf{Effective} \\ \textbf{Mitigations}\end{tabular} \\ 
\hline \hline
NetSpectre~\cite{schwarz2019netspectre} \rule{0pt}{2ex}
	& \cmark 
	& \xmark  
	& \xmark  
	& 1.5 kb/s
    & U
    & Single-level Thread Throttling
	& \cmark 
	& \xmark  
	& \xmark  \\
\hline
TurboCC~\cite{kalmbach2020turbocc} \rule{0pt}{2ex}
	& \xmark  
	& \xmark  
	& \cmark 
	& 61 b/s
    & K
    & Turbo Frequency Change
	& \xmark  
	& \xmark  
	& \xmark  \\
\hline
\tech \rule{0pt}{4.2ex}
	& \cmark 
	& \cmark 
	& \cmark 
	& 3 kb/s
    & U
    & \begin{tabular}{@{}c@{}}Multi-level Thread, SMT,\\ and Core (VR) Throttling \end{tabular}
	& \cmark 
	& \cmark 
	& \cmark \\
\hline
\end{tabular}
}
\caption{\jd{Comparison to state-of-the-art covert channels utilizing throttling effects of current management mechanisms.}}
\label{tab:prior_phi_works1}
\end{center}
\vspace{-10pt}
\end{table*}

%% file: _09_conclusion.tex
\section{Conclusion}

{Based on our rigorous characterization of current management mechanisms in real modern systems, we develop a deep understanding of the underlying mechanisms related to power management for PHIs {(power-hungry instructions)}. In particular, we notice that these mechanisms throttle the execution for a period of time whose length is related to other \agyfour{instructions and} PHIs executing on the system. We exploit these throttling side-effects to develop a set of covert channels, {called \emph{\tech}}. These {high-throughput} covert channels can be established} between two execution contexts running on 1) the same hardware thread, 2) the same simultaneous multi-threaded (SMT) core, and 3) different physical cores.
We implement a proof-of-concept of \agyfour{\tech on} recent Intel processors. We show that \tech can reach {more than} $24\times$ \agyfour{the} channel capacity \agyfour{of} other recent covert channels leveraging power management techniques. We propose multiple practical mitigations to protect \agyfour{against \tech} in modern processors. {We conclude that current management mechanisms in modern processors can lead to malicious information leakage and hope our work paves the way {for} eliminating the confidentiality breaches such processor mechanisms lead to.}